\documentclass[11pt]{JHEP3}
\usepackage{graphicx}
\usepackage{epsfig}
\usepackage{amssymb}
\input epsf.tex

\newcommand{\bi}{\begin{itemize}}
\newcommand{\ei}{\end{itemize}}
\newcommand{\beas}{\begin{eqnarray*}}
\newcommand{\eeas}{\end{eqnarray*}}
\newcommand{\NCS}{N_{\rm CS}}
\newcommand{\nw}{N_{\rm w}}

\newcommand{\mueff}{\mu_{\rm eff}}
\newcommand{\mhinv}{m_H^{-1}}
\newcommand{\ffd}{\Tr [F\tilde F]_{{\rm lat},x}}

\newcommand{\intvecx}{\int d^3 x\,}

\newcommand{\vac}{|0\rangle}

\newcommand{\veck}{{\bf k}}

\newcommand{\vecx}{{\bf x}}       

\newcommand{\vecn}{{\bf n}}
\newcommand{\al}{\alpha}
\newcommand{\bt}{\beta}
\newcommand{\gm}{\gamma}
\newcommand{\dl}{\delta}
\newcommand{\ep}{\epsilon}

\newcommand{\kp}{\kappa}
\newcommand{\lm}{\lambda}
\newcommand{\rh}{\rho}
\newcommand{\sg}{\sigma}
\newcommand{\ta}{\tau}

\newcommand{\ph}{\phi}

\newcommand{\ch}{\chi}

\newcommand{\om}{\omega}

\newcommand{\Gm}{\Gamma}

\newcommand{\hmu}{\hat{\mu}}

\newcommand{\half}{\frac{1}{2}}
\newcommand{\quart}{\frac{1}{4}}
\newcommand{\third}{\frac{1}{3}}
\newcommand{\Tr}{\mbox{Tr}\,}

\newcommand{\dmu}{\partial_{\mu}}

\newcommand{\dnot}{\partial_{0}}

\newcommand{\phd}{\ph^{\dagger}}

\newcommand{\eela}[1]{\label{#1}\end{equation}}
\newcommand{\eeala}[1]{\label{#1}\end{eqnarray}}
\newcommand{\be}{\begin{equation}}
\newcommand{\ee}{\end{equation}}
\newcommand{\bea}{\begin{eqnarray}}
\newcommand{\eea}{\end{eqnarray}}

\newcommand{\hnu}{\hat{\nu}}

\title{Baryon asymmetry from electroweak tachyonic preheating}

\author{Anders Tranberg and Jan Smit\\
Institute for Theoretical Physics, University of Amsterdam, \\
       Valckenierstraat 65, 1018 XE Amsterdam, the Netherlands.\\
}

\keywords{Baryogenesis, CP violation, Out-of-equilibrium field theory, Preheating, Symmetry breaking}

\preprint{ITFA-2003-49}

\abstract
{We consider a scenario in which the baryon asymmetry was created in the early
universe during a cold electroweak transition. The spinodal instability
of the Higgs field caused by a rapid change of sign of its effective
mass-squared  parameter induces tachyonic preheating.
We study the development of Chern-Simons number in this transition by numerical lattice simulations of the SU(2)-Higgs model with an added effective CP-violating term. A net asymmetry is produced, and we study its dependence on the size of CP violation and the ratio of Higgs to W mass.}

\begin{document}


\section{Introduction}
\label{Introduction}
The baryon number asymmetry of the Universe, expressed in terms of the observed ratio of baryon number density ($n_{B}$) to photon number density ($n_{\gamma}$) \cite{Spergel:2003cb}
\bea
\label{ObsAsy}
\frac{n_{B}}{n_{\gamma}}= 6.5^{+0.4}_{-0.3}\times10^{-10}
\eea
is thought to be the result of high energy processes in the very early Universe. These processes need to break charge conjugation symmetry (C), the symmetry under the combined charge conjugation and parity (CP) and conservation of baryon number (B). In addition it is necessary for the processes to take place during a period when the Universe was sufficiently out of thermal equilibrium
\cite{Sakharov:1967dj}.

Quite a few theories of baryogenesis have been proposed that can
explain the order of magnitude of the asymmetry (\ref{ObsAsy}),
see e.g.\ the review \cite{Dine:2003ax}. Most of these are based on
physics beyond the Standard Model (SM) and because of our limited
knowledge of this physics and of the history of the very early universe,
it is hard to falsify a particular proposal. We consider it
therefore important to search for a viable scenario of
baryogenesis within the known physics of the Standard Model.
We should also include its renormalizable extension that includes
right-handed neutrino fields and Dirac and Majorana mass terms ---
which we shall call the Extended Standard Model (ESM) --- to
incorporate the neutrino masses.

In the (E)SM all criteria for baryogenesis are met. Baryon number
is violated through an anomaly \cite{'tHooft:1976up} and the weak
interactions violate C and CP. For three or more generations, CP
violation in the quark sector is possible through the
Cabibbo-Kobayashi-Maskawa (CKM) mixing matrix 
\cite{Cabibbo:1963yz,Kobayashi:1973fv}.
Experimentally, three generations are observed and the measured
CP violation in Kaon and heavy-quark systems is consistent with
the CKM matrix \cite{Hagiwara:2002fs}. There may be more sources
of CP violation in the neutrino sector, thus far hidden in the
neutrino mixing matrix. 
The required non-equilibrium may be found in the decay of heavy
neutrinos falling out of thermal equilibrium at temperatures $T >
10^8$ GeV
\cite{Buchmuller:2003gz} --- the leptogenesis scenario --- or in the
electroweak transition.

\subsection*{Tachyonic electroweak transition}

It is a great challenge to develop a scheme in which
the baryon asymmetry follows from the physics of the (E)SM at an
energy scale of order 100 GeV. At this energy scale the Hubble rate is
only about $10^{-5}$ eV and the required
non-equilibrium dynamics is then to be caused by the electroweak transition.
Such schemes fall under the unifying name of electroweak baryogenesis
\cite{Kuzmin:1985mm,Rubakov:1996vz,Cohen:1993nk}.
Most schemes rely on the electroweak transition being a
sufficiently strong first-order finite-temperature phase transition.
It was found later that the electroweak transition cannot be
first order in the Standard Model \cite{Kajantie:1996kf,Csikor:1998eu},
given the experimental bound on the Higgs mass $m_{H}>114$ GeV \cite{Hagiwara:2002fs}.
This has led to investigations of whether the phase transition can be first
order in theories with an extended Higgs sector, such as the
Minimal Supersymmetric Standard Model \cite{Carena:1996wj,Laine:1998qk}.

Another possibility for the required out-of-equilibrium conditions is
preheating at the electroweak scale after low-scale inflation, either
through resonant preheating \cite{Garcia-Bellido:1999sv,Krauss:1999ng},
or tachyonic preheating and the creation of topological defects
\cite{Krauss:1999ng,Copeland:2001qw}. In these scenarios
the electroweak transition is not a regular finite-temperature transition
but a {\em cold} transition, caused by a hybrid-inflation type
coupling of the Higgs field to an inflaton field. Such electroweak
transitions have been investigated in more detail
\cite{Rajantie:2000nj,Copeland:2002ku}, and the tachyonic
case appears most promising
\cite{Garcia-Bellido:2002aj,Smit:2002yg,Garcia-Bellido:2003wd}.

In such a tachyonic electroweak scenario for
baryogenesis, the baryon number generated during the transition is
given by the anomaly equation
\bea
B(t)=3 \langle N_{\rm CS}(t)-N_{\rm CS}(0)\rangle,
\label{Anomaly}
\eea
where $\NCS$ is the Chern-Simons number in the SU(2) gauge fields
and $B$ is assumed to be negligible (e.g.\ due to inflation)
before the transition at time $t=0$.
A sufficient CP bias is needed to produce the observed asymmetry (\ref{ObsAsy}).

\subsection*{CP violation}

A further hurdle to be overcome is the strength of CP violation.
In the Standard Model this has been estimated as being of order
\cite{Jarlskog:1985ht}
\be
J\,
(y_u-y_c)^2(y_c-y_t)^2(y_t-y_u)^2
(y_d-y_s)^2(y_s-y_b)^2(y_b-y_d)^2
\approx 10^{-23},
\label{dlcpest}
\ee
where $y_u$, \ldots, $y_t$ are the fermion-Higgs Yukawa couplings and
\cite{Hagiwara:2002fs}
\be
J = |{\rm Im} (V_{ij}V_{kl}V_{il}^* V_{kj}^*)| = (3 \pm 0.3) \times 10^{-5}
\label{J}
\ee
is the simplest rephasing-invariant combination of the CKM matrix
$V$. In
\cite{Shaposhnikov:1987tw,Shaposhnikov:1988pf,Rubakov:1996vz}
finite-temperature estimates were given with $y_f \to m_f^2/T^2$
and
$T$ of order of the electroweak transition, $T\approx 100$ GeV, which led to
a magnitude $\approx 10^{-20}$.
Clearly, if these estimates are to be taken seriously,
Standard Model CP-violation is much too weak for baryogenesis, which is
the current lore.

CP violation is a subtle effect based on quantum mechanical interference,
which can be destroyed at high temperatures.
The above finite-temperature estimate can perhaps be justified by
a high-temperature expansion and dimensional reasoning.
However, the zero temperature estimate (\ref{dlcpest}) seems less
reliable to us, since it contradicts the
measured magnitude of CP-violating observables. Factors such as $J$
are hard to avoid in analytical calculations,
but the product of Yukawa couplings in (\ref{dlcpest}) could be compensated
by perturbative energy-denominators or non-perturbative effects
(see e.g.\ \cite{Farrar:1993sp,Farrar:1994hn,Farrar:1994kf}
and \cite{Konstandin:2003dx}).
We found indications of such compensation in a computation of the effective
action obtained by integrating out the fermions \cite{Cosmo03JS}.

One motivation for our work here is the
possibility that {\em at zero temperature} the strength of CP violation
in the SM is given by $J$, i.e.\ without the product of $y$'s.
This is another reason for considering scenarios based on a tachyonic
electroweak transition.
In the ESM with Majorana masses there is less rephasing-invariance and the
corresponding $J' = {\rm Im}(V'_{ij} V^{\prime *}_{ik})|$ could be even larger.

\subsection*{Sphaleron transitions}

It is important that the effective temperature after the transition is low
enough for sphaleron processes to be strongly suppressed,
otherwise $B$ might diffuse to zero again
(see \cite{Bodeker:1999gx,Moore:1998sw} for recent results on the magnitude
of the sphaleron rate).
Because a tachyonic transition occurs at a relatively low energy,
which becomes redistributed
over the many degrees of freedom by the efficiency of
the preheating process, sphaleron wash-out is not expected to
be a problem, as already suggested
in \cite{Garcia-Bellido:1999sv}.
A recent study \cite{Skullerud:2003ki}
of Higgs- and W-particle numbers shortly after
a tachyonic electroweak transition indeed showed low effective
temperatures. However, particle numbers in the important low momentum
modes were found to be still rather large, corresponding to large
chemical potentials \cite{Skullerud:2003ki}, suggesting the possibility
of a sizable effective sphaleron rate after the transition.

\subsection*{This work}

In this paper we report on a study of the asymmetry generated in a
tachyonic electroweak transition
by lattice simulations of the SU(2)-Higgs model
with effective CP violation. The transition is modelled by a rapid
quench and we use a classical approximation that can be justified for
such transitions \cite{Garcia-Bellido:2002aj,Smit:2002yg};
the quantum average is approximated by an average over a classical
ensemble of initial conditions, which is evolved using classical
equations of motion.

The effective CP violation we use is given by a term
\bea
\kappa\, \phi^{\dagger}\phi \Tr[F^{\mu\nu}\tilde{F}_{\mu\nu}]
\label{CPterm}
\eea
in the lagrangian
($\ph$ is the Higgs doublet, $F^{\mu\nu}$ the gauge field strength tensor and
$\tilde{F}_{\mu\nu}$
its dual). The coefficient $\kappa$ parametrizes the CP violation and
has mass dimension $-2$. One may write it as
\bea
\kappa=\frac{3\delta_{CP}}{16\pi^2 M^2}
\label{deltacp}
\eea
where $M$ is some mass scale and $\delta_{CP}$ is dimensionless.

There are two possible interpretations of the term (\ref{CPterm}).
We can see it as a crude representation of the CP violation due to
the fermions in the (E)SM, in which case we choose $M = m_W$. In
the second interpretation the term (\ref{CPterm}) could be the
lowest-dimensional CP-violating operator resulting from
integrating out heavy fermions (with masses beyond the electroweak
scale) in a path integral, in a large class of theories beyond the
Standard Model. In this case we choose $M=1$ TeV. Notwithstanding
the approximate nature of the model, which neglects the dynamics
of the (E)SM fermions as well as the U(1) and SU(3) gauge fields,
it is very interesting to obtain an estimate of the value of
$\dl_{CP}$ that is needed for reproducing the asymmetry (\ref{ObsAsy})
via (\ref{Anomaly}), in either interpretation.

With one exception \cite{Ambjorn:1991pu}, simulations in 3+1
dimensions have not included CP violation. In
\cite{Ambjorn:1991pu} there was a search for an asymmetry induced
by a chemical potential for Chern-Simons number. Technical
difficulties were reported and no asymmetry was seen. 
Simulations in 1+1 dimensions similar to what we
consider here were performed in
\cite{Grigoriev:1992nv,Garcia-Bellido:1999sv} and in more detail for
the case of a tachyonic quench in
\cite{Smit:2002yg}. In
\cite{Smit:2002yg} we showed that the final asymmetry is
proportional to the applied C 
violation (in 1+1 dimensions, C
violation plays the role of CP violation in 3+1 dimensions), with
an interesting dependence on the ratio of Higgs to W masses.

The structure of the rest of this paper is as follows:
In section \ref{Preheating} we review scenarios for tachyonic preheating
after inflation.
In section \ref{SU2} we describe the SU(2)-Higgs model with the
added CP-violating term, and discuss the equations of motion.
Section \ref{Init} is devoted to the classical approximation and the
details of the applied initial conditions. In section \ref{Results}
we present our results. Finally, section \ref{Conc} concludes
with our estimate for the generated baryon asymmetry and
the magnitude of $\dl_{CP}$. Technical details can be found in the Appendix.

Preliminary accounts of this work have been given in
\cite{SEWM02AT,Smit:2003my,Cosmo03AT,Cosmo03JS}.

\section{Tachyonic electroweak preheating}
\label{Preheating}

As mentioned in the Introduction, with regard to the magnitude of
CP violation it is interesting to explore scenarios in which the
electroweak transition took place at zero temperature. Let us
assume that the effective mass parameter
$\mueff^2$ in the Higgs potential
\be
V_{\phi}=V_0
+\mueff^2\,\phi^{\dagger}\phi +\lambda(\phi^{\dagger}\phi)^{2}
\ee
was at first positive and then turned negative (`tachyonic'),
ending up at the current value $-\mu^2$. We recall that $\mu$ is
simply related to the Higgs mass,
$\mu^2 = m_H^2/2$,
and assume for the purpose of discussion that
$\mu \approx 100$ GeV.
We now discuss two realizations for the sign change of $\mueff^2$,
firstly one that is instructive but that we will argue to be not
viable, and secondly the more conventional one proposed in
\cite{Copeland:2001qw}.

An elegant realization for $\mueff^2$ could be a non-minimal
coupling to the gravitational field,
\be
\mueff^2=\xi R -\mu^{2},
\label{xireal}
\ee
with $R$ the scalar curvature. As far as we know, there are no limits known yet
on the parameter $\xi$. An indicative value is $\xi = 1/6$, the conformal
case, so it is natural to assume $\xi \approx 1/6$. As an example, consider
inflation in which $R$ is initially large and positive, such that
$\mueff^2$ is positive. After inflation $R$ goes down and
$\mueff^2$ goes through zero when $R\approx (100\, \mbox{GeV})^2$.
The transition will take place somewhat later when the Hubble rate,
which is generically of order $R^{1/2}$, has dropped sufficently
for the Hubble damping term in the Higgs-field e.o.m.\
($3 H \dot \ph$) to become subdominant, say $H \approx 1 - 10$ GeV.
For $H \approx  1$ GeV, the energy density is still very large,
$\rh = 3 H^2 m_{\rm P}^2\approx (10^9 \mbox{GeV})^4$ 
(we use $m_{\rm P} = (8\pi G)^{-1/2}$), far above the
electroweak scale of about $(100\, \mbox{GeV})^4$.  After
the transition the inflaton still dominates the energy and
it should have released this into the SM degrees of freedom before
big bang nucleosynthesis (BBN) can take over.

In the conventional scenario (see e.g.\ \cite{KT}, section 8.3),
the universe expands in matter-dominated fashion when the inflaton
($\sg$) has fallen out of slow roll and is oscillating
in the minimum of its potential, while decaying into the SM degrees 
of freedom (`radiation'). The
inflaton decay rate $\Gm_\sg$ should be sufficiently slow,
otherwise the maximum temperature in radiation \cite{KT},
\be
T_{\rm max} \simeq g_*^{-1/4} M_{\rm infl}^{1/2} (\Gm_\sg m_{\rm P})^{1/4},
\ee
might rise above the electroweak transition temperature $T_c \approx 100$ GeV,
and the sphaleron rate would wash out the generated asymmetry.
Here $g_*$ is the usual effective number of d.o.f\
and $M_{\rm infl}$ is the energy scale of
inflation, which by the previous argument is larger than $\approx 10^9$ GeV.
Requiring $T_{\rm max} < 100$ GeV leads to $\Gm_\sg < 10^{-27}$ GeV.
Such an extremely small decay rate is perhaps
not unnatural if the inflaton resides
in gravitational degrees of freedom,
and $(\xi - 1/6)^2$ is sufficiently small.
However, it also leads to a reheating temperature \cite{KT}
\be
T_{\rm rh} \approx  g_*^{-1/4} (\Gm_\sg m_{\rm P})^{1/2}
\ee
that is
smaller than $10^{-2}$ MeV, in conflict with BBN. Moreover, the entropy
in a comoving volume produced by the decaying inflaton,
$S= s a^3$ ($s$ is the entropy density and $a$ the scale factor),
increases $\propto a^{15/8}$ \cite{KT}, whereas the corresponding baryon number
$B = n_B a^3$ is essentially conserved for $T<100$ GeV. Then the crucial ratio
$n_B/s$ suffers dilution by a huge factor
$(a_1/a_2)^{15/8} = (t_1/t_2)^{5/4} \approx 10^{-34}$,
where $t_1\approx m_{\rm P}/M_{\rm infl}^2$ is the time at the end of
inflation and $t_2 \approx 1/\Gm_\sg$
the time the inflaton has decayed.

The numbers look better, but not good enough, if
the inflaton potential does not have quadratic minimum but simply falls away
such that the kinetic energy dominates (`kinetion'),
as in quintessential inflation \cite{Peebles:1998qn}.
In this case the energy density falls 
faster than the radiation in SM particles,
$\rh_\sg \propto a^{-6}$.
However, going through a similar analysis leads to practically the same
$T_{\rm rh}$, an entropy growth
$S \propto a^{3/4}$ and a dilution factor $\approx 10^{-8}$.
In case $\Gm_\sg$ is negligible, the entropy would be conserved and
$n_B/s$ would be constant. However, it then takes too long for
the energy density in SM d.o.f., $\rh_{\rm SM} \propto a^{-4}$, to
become comparable to $\rh_\sg$. This would happen at a temperature
much too low for BBN:
$T/T_1 = a_1/a = (\rh_{\rm SM}/\rh_\sg)_1^{1/2} \approx 10^{-14}$,
or $T \approx 10^{-9}$ MeV.

These problems are avoided if the energy scales of inflation and
radiation are comparable, specifically low-scale inflation with
$M_{\rm infl}\approx 100$ GeV. For this the realization (\ref{xireal}) does
not work (unless $\xi > 10^{26}$!).
In \cite{Garcia-Bellido:1999sv,Krauss:1999ng} a hybrid inflation mechanism
was proposed in which the Higgs field is coupled to the inflaton,
\be
\mueff^{2}=\lm_{\sg\ph}\,\sigma^2 -\mu^{2},
\ee
and
$\sigma$ was assumed to roll from large values towards $\sigma=0$.
Considering the effect of radiative corrections of the Higgs to
the inflaton led \cite{Copeland:2001qw}
to the conclusion that inverted hybrid inflation is a
better option (see also \cite{German:2001tz}),
\be
\mueff^{2}=\mu_{\ph}^{2} - \lm_{\sg\ph}\,\sigma^2,
\ee
where now $\sigma$ runs from $0$ to
$[(\mu^{2}+\mu_{\ph}^{2})/\lm_{\sg\ph}]^{1/2}$.
An explicit example of
the inflaton-Higgs potential is given in \cite{Copeland:2001qw}:
\be
V(\sg,\ph) = V_0 - \frac{1}{5}\, \lm_5 \sg^5 + \frac{1}{6} \kp_6 \sg^6
+ \half (\mu_{\ph}^{2} - \lm_{\sg\ph}\,\sigma^2)\ph^2 + \quart \lm \ph^4
\label{Copeeamod}
\ee
(for simplicity for one real $\ph$),
with
$\lm_5 = 7.3\times 10^{-5}\, \mbox{GeV}^{-1}$,
$\kp_6 = 2.4 \times 10^{-7}\, \mbox{GeV}^{-2}$,
$\mu_\ph^2 = 1000\, \mbox{GeV}^2$,
$\lm_{\sg\ph} = 0.04$,
and $V_0$ such that $V=0$ in its minimum, $V_0 \simeq (86\, \,\mbox{GeV})^4$.

Choosing $\lm = 1/9$ leads to the vacuum expectation values
$\langle \ph\rangle = 237$ GeV (reasonably close to the correct
 246 GeV),
$\langle \sg\rangle  = 426$ GeV,
and $\mu = 79$ GeV.
The inflaton-Higgs coupling $\lm_{\sg\ph}$ is quite strong, which
causes substantial mixing in the minimum
of the potential. The eigenvalues of the mass matrix are given by
$(74\, \mbox{GeV})^2$ and $(147\, \mbox{GeV})^2$,
differing considerably from the diagonal elements
$\partial^2 V/\partial\sg^2 = (121\,\mbox{GeV})^2$ and
$\partial^2 V/\partial\ph^2 = (111\,\mbox{GeV})^2$,
so one may wonder if the SM Higgs physics is not too much
affected.

Another question that needs further investigation is the fact that
in this model the inflaton decay width is much larger than the Hubble rate.
At this low-scale inflation
the Hubble rate is only $H\approx 10^{-5}$ eV, whereas one expects
$\Gm_\sg\approx 1$ GeV. In the regime $\Gm_\sg >> H$, {\em warm inflation}
has been advocated \cite{Berera:2003kg}, which puts into question our assumption
of a cold universe after inflation.

However, the problems with the previous realization (\ref{xireal})
are avoided. After the transition the inflaton decays rapidly into the
SM d.o.f.\ via the mixing with the Higgs. Neglecting the tiny
Hubble rate and using energy conservation, the `reheat'
temperature is approximately given by
\be
T = \left(\frac{30}{\pi^2 g_*}\right)^{1/4} V_0^{1/4}.
\label{Testimate}
\ee
With $g_* = 86.25$ for the effective number of d.o.f.\
below the W mass (leptons, quarks, gluons, photons), and a Higgs mass of,
say, 160 GeV, this gives
$T \simeq 51$ GeV, well above the BBN and QCD transition temperatures.

Because of the uncertainties in the specific realization of the
assumed tachyonic electroweak transition, we model the transition
by an instantaneous  quench:
\bea
\mueff^2&=&+\mu^{2},\;\;\;\; t<0,
\nonumber\\
&=&-\mu^{2},\;\;\;\; t>0.
\eea
This is a limiting case of the transition in a finite time treated in
\cite{Asaka:2001ez,Garcia-Bellido:2003wd}.
In this way we `shield' the SM from the uncertainties of the
inflaton and presumably, the quench will generate a maximal baryon
asymmetry for a given amount of CP violation. Requiring zero
vacuum energy gives
\be
V_0 = \mu^4/4\lm,
\ee
with a `reheat' temperature similar to the model (\ref{Copeeamod}).

The initial state for the Higgs field is $\langle\phi\rangle=0$
with $\mueff^{2}>0$. When $\mueff^{2}<0$ the Higgs field suffers
the spinodal instability:
$\langle \phi\rangle$ does not change, but its low momentum modes
grow exponentially 
\cite{Smit:2002yg}. 
Soon
the quartic term in the Higgs potential kicks in and eventually
with the other SM couplings the fields will thermalize in the
broken phase minimum. Initially there is rapid
effective-thermalization \cite{Skullerud:2003ki}, called `tachyonic
preheating'
\cite{Felder:2000hj}.

\section{The SU(2)-Higgs model}
\label{SU2}
To study the baryon asymmetry emerging after the electroweak transition
we used the SU(2) Higgs model
with effective CP violation, given by the action
\bea
S= -\int d^{4}x\left[\frac{1}{2g^{2}}\Tr F_{\mu \nu}F^{\mu \nu}
+(D_{\mu}\phi)^{\dagger}D^{\mu}\phi-\mu^{2}\phi^{\dagger}\phi
+\lambda(\phi^{\dagger}\phi)^{2}+V_{0} \right.
\nonumber\\
\left.+\,\kappa\phi^{\dagger}\phi\, \Tr F^{\mu\nu}\tilde{F}_{\mu\nu}\right]
\label{cont_act}
\eea
Here
$F_{\mu\nu}=\partial_{\mu}A_{\nu}-\partial_{\nu}A_{\mu}-i[A_{\mu},A_{\nu}]$,
$D_{\mu}\phi=(\partial_{\mu}-iA_{\mu})\phi$,
with $\phi$ the Higgs doublet,
$\tilde{F}_{\mu\nu}=\epsilon_{\mu\nu\rho\sigma}F^{\rho\sigma}/2$,
our metric is $(-1,1,1,1)$ and $\ep_{0123} = +1$. As usual
$A_{\mu}=A_{\mu}^a
\ta^a/2$ with $\ta^a$, $a=1,2,3$, the Pauli matrices. The vacuum
expectation value of the Higgs field is $|\langle\ph\rangle| =
v/\sqrt{2}$, with $v=\mu/\sqrt{\lambda}$, and the Higgs and W
masses are given by
$m_H = \sqrt{2}\,\mu = \sqrt{2\lm}\, v$, $m_W = g v/2$;
$V_0 = \mu^4/4\lm$.
The magnitude of the effective CP violation is parametrized by $\kp$.
For later reference, we define the dimensionless
\be
k = 16\pi^2 \kappa m_{W}^{2}.
\ee
The equation of motion for the Higgs doublet is 
\be
\left(D_{\mu}D^{\mu} + \mu^2 -2\lm\phd\ph-\kp \Tr F^{\mu\nu} \tilde F_{\mu\nu}
\right) \ph = 0.
\ee
For the gauge fields the equations of motion are given by
\be
D_0\left(\frac{1}{g^2}\,E_{k}^a -2\kp \phd\ph B_k^a\right)
-\ep_{klm}D_l\left(\frac{1}{g^2}\,B_m^a + 2\kp \phd\ph
E_{m}^a\right) + j^a_k = 0,
\label{eom1}
\ee
where $E^a_k = F^a_{k0}$ and $B_k^a = \ep_{klm} F^a_{lm}/2$ are
the SU(2) electric and magnetic fields,
$D_k$ is the covariant derivative in the adjoint representation, e.g.\
\be
D_{\mu} B_k^a = \dmu B_k^a + \ep_{abc} A_{\mu}^b B_k^c,
\ee
and $j^a_{\mu}$ is the Higgs contribution to the SU(2) current
\be
j^a_{\mu} = i(D_{\mu}\ph)^{\dagger} \frac{\ta^a}{2}\ph -i
\phd\frac{\ta^a}{2}D_{\mu}\ph.
\ee
In addition, the Gauss constraints have to be satisfied,
\be
D_k \left(\frac{1}{g^2}\, E_{k}^a -2 \kp \phd\ph B_k^a\right)  +
j^a_0 = 0.
\label{gauss1}
\ee
If these conditions hold at one time, they hold at all times as a
consequence of the equations of motion (\ref{eom1}).

If $\phd\ph$ were constant, the CP-violating terms are ineffective, since
then the $\kp$-term in the action is the integral of a total derivative,
\be
\Tr F_{\mu\nu}\tilde F^{\mu\nu} = 16\pi^2 \dmu j^{\mu}_{\rm CS},
\ee
where $j^{\mu}_{\rm CS}$ is the Chern-Simons current
\be
j^{\kp}_{\rm CS} = \frac{1}{32\pi^2}\, \ep^{\kp\lm\mu\nu}
\left(A_{\lm}^a F_{\mu\nu}^a -
\third \ep_{abc} A^a_{\lm} A^b_{\mu} A^c_{\nu}\right).
\label{CScurrent}
\ee
Similarly, for constant $\phd\ph$, the $\kp$ terms drop out of the
field equations because of the Bianchi identities
$\ep^{\kp\lm\mu\nu}D_{\lm} F^a_{\mu\nu} = 0$. Hence, an
alternative version of (\ref{eom1}) and (\ref{gauss1}) is
\bea
0&=& D_0 E^a_k -2g^2\kp B^a_k \dnot(\phd\ph) -\ep_{klm} D_l B^a_m
-2g^2\kp \ep_{klm} E^a_m \partial_l(\phd\ph) + g^2 j^a_k,
\label{eom2}\\
0&=& D_k E^a_k -2g^2\kp B^a_k \dnot(\phd\ph) + g^2 j^a_0,
\label{gauss2}
\eea
respectively.

\section{Initial conditions and classical approximation}
\label{Init}
We consider an initial state where the Universe is at zero temperature,
with the Higgs field expectation value at zero.
In the vacuum, there are quantum fluctuations, and we would like to use
those to seed the Higgs symmetry breaking. Several ways have been suggested
(see for instance \cite{Rajantie:2000nj,Smit:2001qa}).
We follow the line described in \cite{Smit:2002yg}
(see also \cite{Garcia-Bellido:2002aj} for the more realistic case of a
rapid but not instantaneous quench),
by solving the quantum evolution
in the limit of zero coupling, $\lambda=0$.
We also neglect the interactions with the gauge fields for the moment.
In this limit, the Higgs potential after the quench at $t=0$ is just an
inverted parabola
\be
V(\phi)=V_{0}-\mu^{2}\phi^{\dagger}\phi
= V_0 - \half\, \mu^2 \ph_{\al}\ph_{\al}
\ee
where the $\ph_{\al}$, $\al = 1,2,3,4$, are four real fields
representing the complex Higgs doublet.
For the moment we consider just one of those real components.
It is straightforward to solve the operator equations of motion for $t >0$
with the initial condition that the field is free with mass $\mu$ for times
$t <0$.
In terms of Fourier modes
\be
\phi_{\veck} =\intvecx\frac{e^{-i\veck\cdot\vecx}}{\sqrt{L^3}}\, \ph(\vecx)
\ee
in a periodic volume $L^3$, one then finds that modes with $k <
\mu$ are unstable and grow exponentially. For $\sqrt{\mu^2-k^2}\,
t \gg 1$,
\be
\ph_{\veck} \propto \exp\left(\sqrt{\mu^2-k^2}\, t\right),
\ee
and the particle numbers of the unstable modes in the initial
state $\vac$ (the vacuum for $t<0$) also grow exponentially. So
these dominating modes become classical and indeed, the
expectation value of generic products of field operators can be
reproduced by a classical gaussian distribution
\cite{Smit:2002yg},
\bea
P(\xi)&\propto&
\exp\left[-\frac{1}{2}\sum_{|\veck|<\mu}\left(
\frac{|\xi^{+}_{\veck}|^{2}}{n_{k}+1/2+\tilde{n}_{k}}+
\frac{|\xi^{-}_{\veck}|^{2}}{n_{k}+1/2-\tilde{n}_{k}}\right)
\right],
\label{Pinit}
\\
\xi_{\veck}^{\pm}&=&
\frac{1}{\sqrt{2\omega_{k}}}\left(\omega_{k}\phi_{\veck}\pm \pi_{\veck}\right),
\label{xidef}
\eea
where $\pi$ is the canonical conjugate to $\ph$.
The particle numbers $n_k$, $\tilde n_k$ and frequencies $\om_k$
are defined in terms of the two-point functions
\bea
\langle\pi_{\veck}\pi^{\dagger}_{\veck}\rangle&=&
\left(n_{k}+\frac{1}{2}\right)\omega_{k},\\
\langle\phi_{\veck}\phi^{\dagger}_{\veck}\rangle&=&
\left(n_{k}+\frac{1}{2}\right)\frac{1}{\omega_{k}},\\
\langle\pi_{\veck}\phi^{\dagger}_{\veck}\rangle
&=& \langle\pi_{\veck}\phi^{\dagger}_{\veck}\rangle+i
= \tilde{n_{k}}+\frac{i}{2},
\eea
which can be calculated to be \cite{Smit:2002yg}
\bea
\langle\phi_{\veck}\phi^{\dagger}_{\veck}\rangle &=&
\frac{1}{2\omega_{k}^{+}}
\left[1+\left(\frac{\omega_{k}^{+2}}{\omega_{k}^{-2}}-1\right)
\sin^2(\om_k^- t)\right],
\\
\langle\pi_{\veck}\pi^{\dagger}_{\veck}\rangle&=&
\frac{\omega_{k}^{- 2}}{2\omega_{k}^{+}} \left[1+
\left(\frac{\omega_{k}^{+2}}{\omega_{k}^{-2}}-1\right)
\cos^2(\om_k^- t)\right],
\\
\langle\pi_{\veck}\phi^{\dagger}_{\veck}\rangle &=&
\frac{\omega_{k}^{-}}{4\omega_{k}^{+}}
\left(\frac{\omega_{k}^{+2}}{\omega_{k}^{-2}}-1\right)
\sin(2\om_k^- t) + \frac{i}{2},
\\
\omega_{k}^{\pm} &=& \sqrt{\pm\mu^{2}+k^{2}}.
\eea
Explicit expressions for $n_k$, $\tilde n_k$ and $\om_k$ are easily obtained,
which show that for $|\om_k^-|t > 1$,
$n_k + 1/2 + \tilde n_k$ grows exponentially, whereas the difference
$n_k + 1/2 - \tilde n_k$ rapidly approaches zero
(see figure 2 in \cite{Smit:2002yg}).
This means that the typical $\xi_\veck^+$ grows exponentially and
the typical $\xi_\veck^- \to 0$. In terms of $\ph_\veck$ and $\pi_\veck$
the distribution gets squeezed.

It therefore makes sense to treat the dynamics classically,
as soon as the particle numbers are large, $n_{k}\gg 1/2$.
This is the case for sufficiently long ``roll-off'' times,
keeping in mind that the quadratic approximation will break
down when the interaction terms becomes non-negligible.
For a given choice of ``roll-off'' time and thus $n_{k}$,
$\tilde{n}_{k}$ and $\omega_{k}$, we reproduce generic quantum correlators
of the initial conditions by an ensemble of random classical initial field
configurations. These can then be evolved using the classical equations of
motion. In this classical approximation the quantum averages are
replaced by averages over the initial ensemble.

The classical evolution can be carried out on a computer,
including fully the non-linear interactions. This assumes of
course that the particle numbers in the gauge fields will also
grow large, such that they can also be treated classically. A
large growth of gauge field modes could be seen {\em a
posteriori}. In \cite{Skullerud:2003ki}, we found that the Higgs
occupation numbers for the unstable modes indeed become very large
($n_{0}\approx 100$) during the instability. In addition, the gauge
fields also acquire large occupation numbers, thereby supporting
the classical approximation for the whole system, at least for the
time scales under consideration here. The large particle numbers
in the gauge field modes suggest that our major observable to be
computed, $\langle \NCS(t) - \NCS(0)\rangle$, can also be obtained
reasonably accurately in the classical approximation.

The Gauss constraints (\ref{gauss1}) need to be imposed on the
initial conditions. We do this as follows. As long as the
free-field approximation is valid, the gauge fields cannot feel
the exponential growth of the Higgs fields, so we initialize
$A_\mu^a =0$. Then the covariant derivative in (\ref{gauss1}) reduces to an
ordinary derivative and it is straightforward to solve for the
electric field components, $E^a_k$, given the Higgs charge
densities $j^{0a}$ drawn from the distribution (\ref{Pinit}).
However, before doing so one constraint has to be imposed first:
in finite volume the global charge vanishes,
\be
\intvecx j^a_0 = \intvecx \partial_k F^a_{0k}/g^2  = 0.
\ee
We thus modify the distribution (\ref{Pinit}),
\be
P(\xi) \to P(\xi)\, \delta({\bf G}),
\label{Pinit2}
\ee
where
$\delta({\bf G})$ encodes this global Gauss constraint.
This makes the distribution not quite Gaussian.
For details see appendix \ref{Gauss1}.

We will apply two types of initial conditions, dubbed `Just-a-half'
and `Thermal'. For additional details on these schemes, see \cite{Smit:2002yg}.

\subsection{Just-the-half distribution}
\label{half}
In the quadratic approximation, classical and quantum evolution is
identical, and we can therefore choose to sample the initial
distribution (\ref{Pinit}, \ref{Pinit2}) at ``roll-off'' time
zero, when the
$n$'s and $\om$'s are simply given by
\be
n_k = \tilde n_k = 0,
\;\;\;\;
\om_k = \sqrt{\mu^2 + k^2}.
\ee
leaving just the 1/2 in the denominators in (\ref{Pinit}). We only
initialize momentum modes that are unstable and that will acquire
large occupation numbers under the spinodal instability, so
$|\veck|<\mu$. This also avoids initializing modes close to the
lattice cut-off scale
\cite{Smit:2001qa}.
We name these initial conditions: Just-the-half.

\subsection{Thermal distribution}
\label{BE}
A different condition is obtained from an initial state that is
thermal but at a low temperature $T\ll m_{H}$. We consider this possibility
to get some indication of the sensitivity of the results to the
initial conditions.
The initial particle numbers of the Higgs fields are chosen
to be Bose-Einstein distributed:
\bea
n_k = \left(\exp{\left(\omega_{k}/T\right)}-1\right)^{-1},
\;\;\;\;
\tilde n_k = 0,
\;\;\;\;
\omega_{k}=\sqrt{\mu^{2}+k^{2}},
\eea
We use $T/m_{H}=1/10$.
In this case we initialize all the modes, not only the unstable ones,
but of course, only the unstable ones will grow.
In practice the large-momentum tail is so suppressed that it should make little
difference to introduce the cut-off $k<\mu$.
We name these initial conditions: Thermal.

\section{Numerical simulation}
\label{Results}
For the numerical simulation we translated the action to a lattice
in (Lorentzian) space-time according to the usual method of
lattice gauge theory, from which the discretized field equations
follow in the standard fashion. The Gauss constraint
(\ref{gauss1}) is compatible with the discretization and since the
initial $A_{0}
= 0$, it stays zero and we are using temporal gauge. Because of
the CP-violating terms the numerical algorithm for the equations
of motion turns out to be `implicit', causing it to be rather
computer-expensive. Lattice details are given in the appendix.

We evolved ensembles of 45 initial conditions in time using
classical dynamics. In two cases (Thermal initial conditions,
$k\equiv 16\pi^2 \kp m_W^2=0$ and $3$) we averaged over larger
ensembles of about 300 configurations. We used volumes
$(Lm_{H})^{3}=21^{3}$ with periodic boundary conditions, and a
spatial lattice spacing
$am_{H}=0.35$, so with $60^{3}$ lattices. This is a compromise
between computer time and capacity and the need to have enough
unstable modes on the lattice. Here, we have about 50 such modes.
Also, we found that with this choice of lattice spacing
discretization errors in the Chern-Simons number and Higgs-winding
number (see below) were small enough for our purpose.

\subsection{Observables}
The spatial average of $\phd\ph$,
\be
\overline{\phd\ph} \equiv \frac{1}{L^3}\intvecx \phd\ph,
\ee
is a good indicator for the development of the instability in
time. Our most important observable is the Chern-Simons number
\be
\NCS = \intvecx j^0_{\rm CS},
\ee
since the change in time of its expectation value determines the
baryon asymmetry through eq.\ (\ref{Anomaly}),
\be
B(t)=3 \langle N_{\rm CS}(t)-N_{\rm CS}(0)\rangle
= 3 \int_0^t dx^0\intvecx \langle \dmu j^{\mu}_{\rm CS}\rangle
=\frac{3}{8\pi^2}\int_0^t dx^0\intvecx
\langle E^a_k B^a_k \rangle.
\label{Anomaly2}
\ee
Another interesting observable is the winding
number in the Higgs field
\bea
\nw &=& \frac{1}{24\pi^2} \intvecx \ep_{klm} \Tr\left(
\partial_k V V^{\dagger}\partial_l V V^{\dagger}\partial_m V V^{\dagger}
\right),
\\
V &=& \frac{(i\ta^2 \ph^*,\ph)}{\phd\ph} \in \mbox{SU(2)}.
\label{Vdef}
\eea
For configurations near equilibrium with low energy in the
covariant Higgs derivatives, the gauge field is not far from being
pure-gauge,
$A_{\mu}\simeq -i\dmu V V^{\dagger}$,
and then $\NCS\simeq \nw$.

Observables such as the Chern-Simons number and the winding number
are notoriously difficult to implement on the lattice in the
quantum theory, but they turn out to be manageable in our
classical approximation because the high-momentum modes are
suppressed in the initial conditions.

\subsection{Single trajectories}
\label{singletraj}

\FIGURE{\epsfig{file=singletraj_k0_th_1.eps,width=10cm,clip}
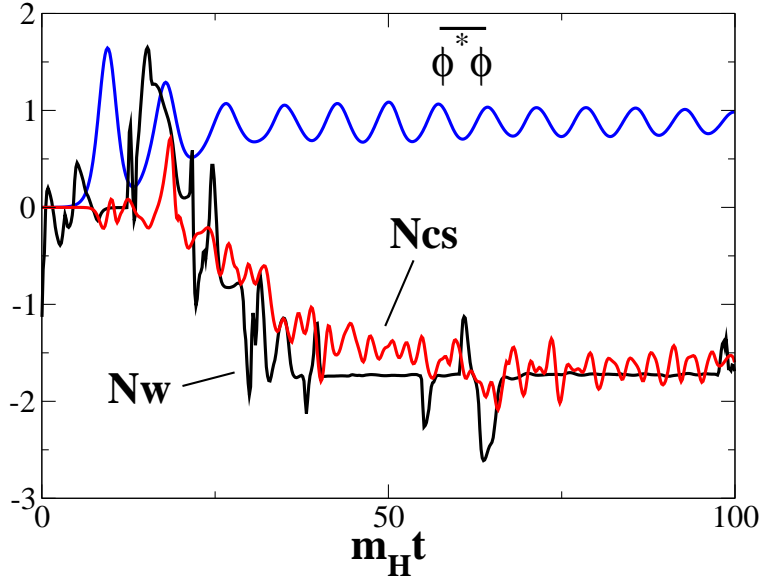
\caption{Volume-averaged Higgs field
$2 \overline{\phd\ph}/v^2$, Chern-Simons number and winding number,
for a typical trajectory with Thermal initial conditions,
$m_{H}/m_{W}=1$, and no CP violation, $k=0$.}
\label{singletraj1}}
\FIGURE{\epsfig{file=singletraj_k0and3_th_1.eps,width=10cm,clip}
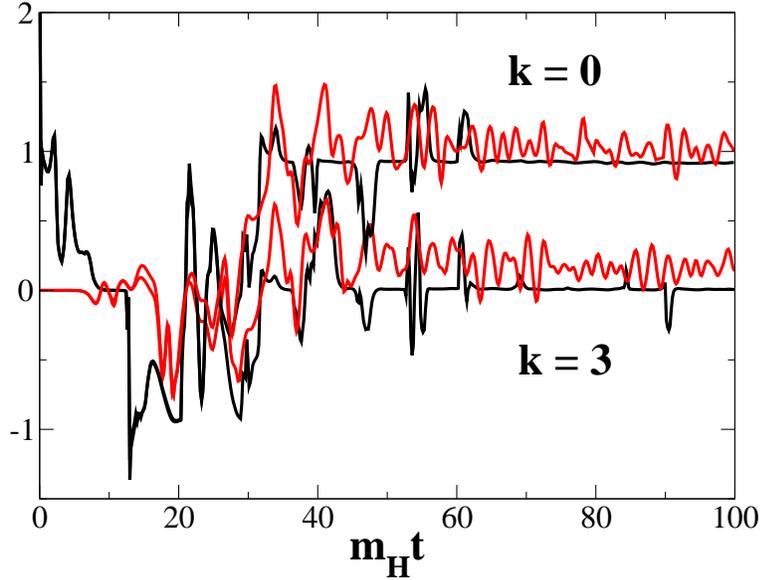
\caption{Example of $\NCS$ and $\nw$ for two trajectories with the same initial
conditions, without ($k=0$) and with ($k=3$) CP violation. Thermal
initial conditions, $m_{H}/m_{W}=1$.}
\label{singletraj2}}

We evolved the initial configurations to time
$t=100\, \mhinv$, 
and computed the volume-averaged Higgs field
$\overline{\phd\ph}$, the Chern-Simons number
$N_{\rm CS}$ and the Higgs winding number $N_{\rm w}$. The
initial Chern-Simons number was set to zero, so we really computed
its change through eq.\ (\ref{Anomaly2}), which is gauge
invariant.

Fig.\
\ref{singletraj1} shows the observables versus time
from a single such trajectory in configuration space. We see that
$\overline{\phd\ph}$ ``falls off the hill into its broken phase
minimum'', where it performs a damped oscillation as the energy
becomes slowly distributed over the higher momentum modes.
Meanwhile, the Chern-Simons number and winding number bounce
around, until they settle near the same integer value (for
$\nw$ the difference with the integer is a finite-lattice spacing
effect). The winding number appears to settle first and then the
Chern-Simons number approaches it somewhat later.

In Fig.\ \ref{singletraj2} we show a particular initial
configuration evolved with and without CP violation. In this case,
the final Chern-Simons number and winding number are shifted
approximately by an integer (1) under the influence of
CP violation.
It is these shifts that will give us a net Chern-Simons number
asymmetry.
Due to the chaotic nature of the equations of motion the shifts
cannot be predicted from the magnitude of the CP violation (for
example, in figure \ref{singletraj2} the final $\NCS \simeq 1$
for $k=0$ and $\simeq 0$ for $k=3$). However, on the average,
more shifts happen to one side than to the other.

\FIGURE{\epsfig{file=histogram_k0.eps,width=10cm,clip}
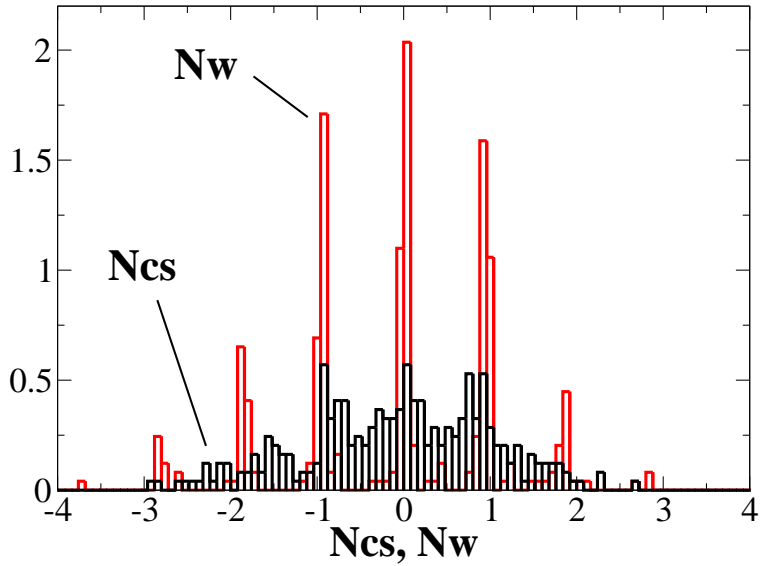
\caption{The distribution of final $N_{\rm CS}$ and $N_{\rm w}$;
Thermal initial conditions, $m_{H}/m_{W}=1$, no CP violation
($k=0$), 303 configurations.}
\label{singletraj3}}

\FIGURE{\epsfig{file=histogram_k3.eps,width=10cm,clip}
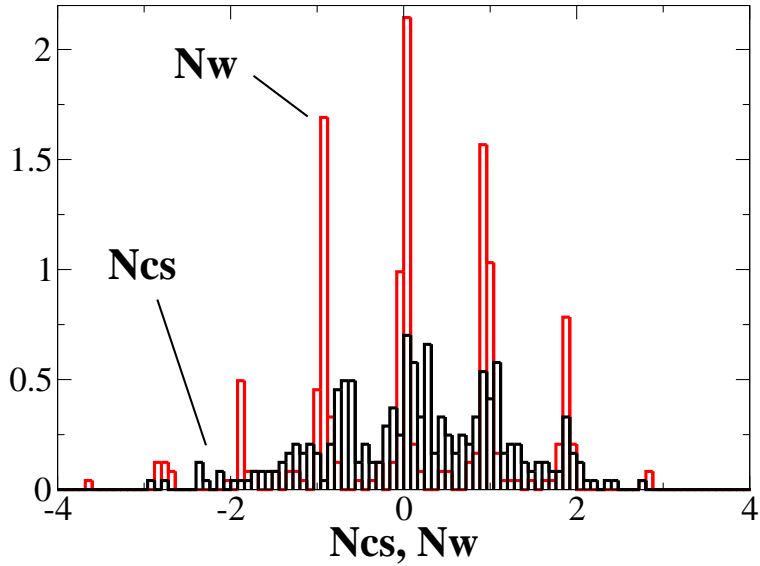
\caption{As in figure \ref{singletraj3} with CP violation, $k=3$.}
\label{singletraj4}}

Fig.\ \ref{singletraj3} shows the distribution of final
Chern-Simons number and winding number. Although both observables
cluster around integers, the winding number is more peaked. The
relatively short evolution time ($m_{H}t_{\rm final}=100$) does
not allow Chern-Simons numbers to settle completely. Furthermore,
the system has a non-zero effective temperature, so the
Chern-Simons numbers need not be integer. We have checked for a
few configurations, that the Chern-Simons number settles
eventually, sometimes as late as
$m_{H}t=500$. For most cases the Chern-Simons number is stuck 
already at time $m_{H}t=100$.

Fig.\ \ref{singletraj4} shows the final distributions when adding
CP violation. Note that the initial configurations in Figs.\
\ref{singletraj3} and \ref{singletraj4} are exactly the same.
It is not obvious that anything has changed and to
spot the asymmetry we have to form ensemble-averaged quantities.

\subsection{Ensemble averages}
\label{averages}

\FIGURE{\epsfig{file=avtraj_k16_h_1.eps,width=10cm,clip}
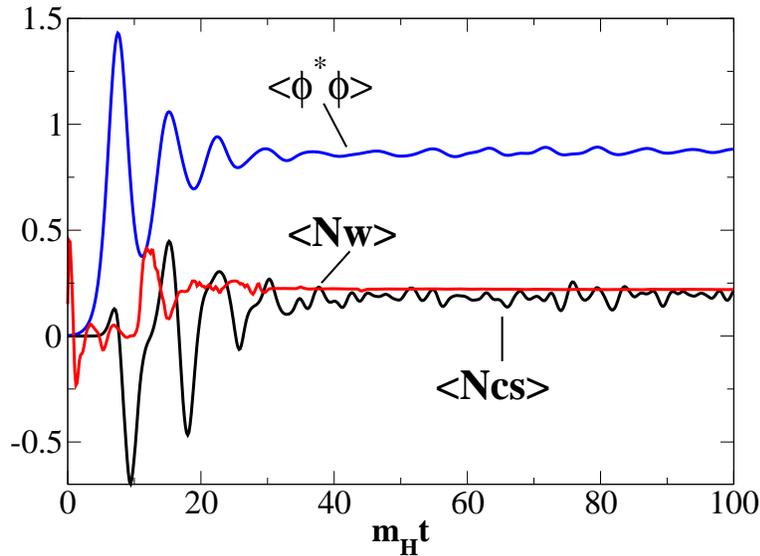
\caption{Averaged observables:
$2\langle\overline{\phi^{\dagger}\phi}\rangle/v^2$,
$\langle\NCS\rangle$ and $\langle\nw\rangle$ versus time for
$m_{H}/m_{W}=1$, Just-a-half initial conditions and
$k=8$.}
\label{averages1}}
In Fig.\ \ref{averages1} we plot the ensemble averaged Higgs
expectation value, Chern-Simons number and winding number vs.\
time. We see that the average Higgs field damps more strongly. The Chern-Simons
number oscillates with a large amplitude before settling down. The
oscillations seem to be driven by the oscillations in the Higgs
field. At the end, both observables settle at non-zero values.
\FIGURE{\epsfig{file=avtraj_allk_Th_1.eps,width=10cm,clip}
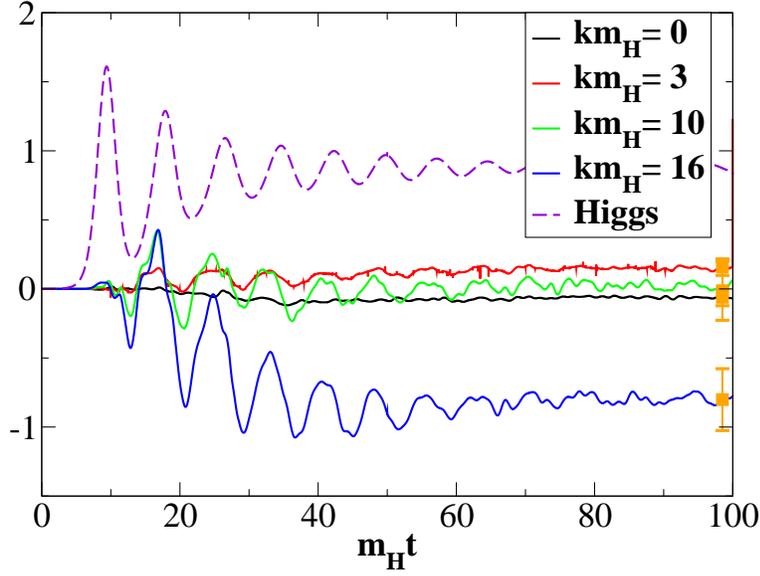
\caption{Expectation values
$2\langle\overline{\phi^{\dagger}\phi}\rangle/v^2$
and $\langle\NCS\rangle$ for different $k$;
$m_{H}/m_{W}=1$ and Thermal initial conditions.}
\label{averages3}}
Fig.\ \ref{averages3} shows the average Chern-Simons numbers for
different values of $k$. Notice that the oscillations have larger
amplitude with larger $k$.

\FIGURE{\epsfig{file=allkandm.eps,width=13cm,clip}
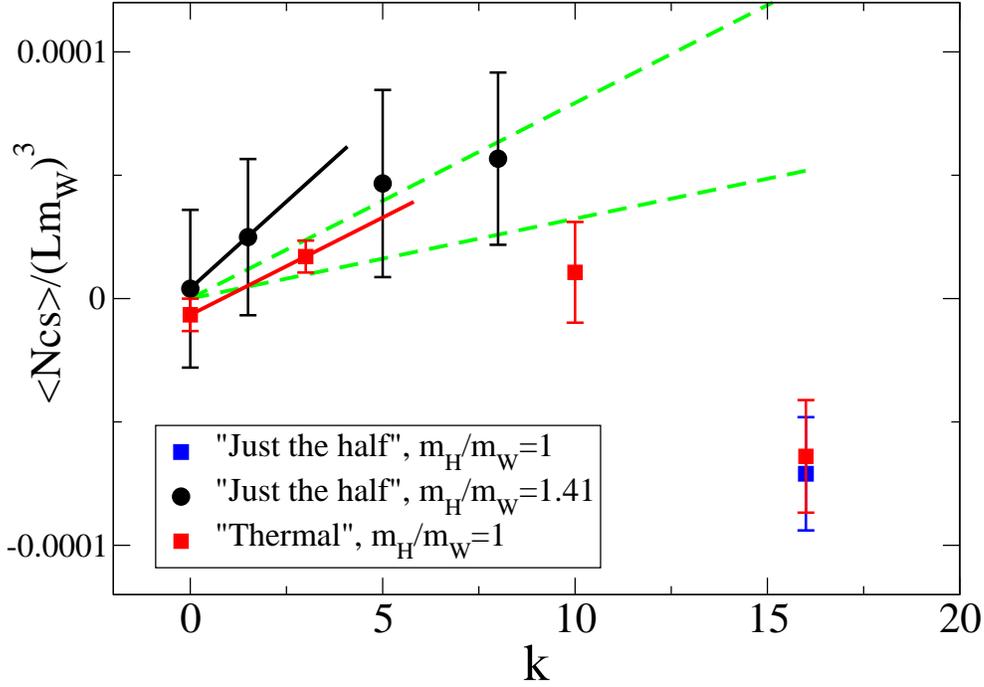
\caption{Results for the Chern-Simons density all $k$,
$m_{H}/m_{W}$ and initial conditions.
The full lines represent linear fits through the two points with
lowest $k$: $k=0,3$ ($m_H = m_W$) and $k=0, 1.5$ ($m_H =
\sqrt{2}\, m_W$). The dashed lines are linear fits through the origin,
ignoring the data at $k=0$, and also at $k=16$ for the case $m_H = m_W$.}
\label{averages5}}
Fig.\ \ref{averages5} sums up our results and the dependencies on
the parameters. We keep $m_{W}$ fixed and plot 
$\langle N_{\rm CS}\rangle/(Lm_{W})^{3}$ vs.\ $k$
at $m_H t = 100$. The dependence on
initial conditions appears to be weak. There is a hint of a
dependence on mass ratio $m_{H}/m_{W}$.

The dependence on
$k$ is not linear for the Thermal, $m_H/m_W=1$, case.
We have observed similar non-linear behavior for large
$\kp$ (large $C$-violation) in the 1+1 D Abelian Higgs model,
but were able to establish linearity at smaller $\kp$. Here a
linear regime presumably also exists at small $k$, but it is
unclear where it ends. We lack sufficient statistics to draw a
conclusion.
Still, a linear fit through the origin is
consistent with the data for $k \leq 10$
(dashed lines), for the $m_H=\sqrt{2}\, m_W$
case ($\chi^2/{\rm d.o.f}=0.25$) and less so for $m_H=m_W$
($\chi^2/{\rm d.o.f}=2.4$),
\bea
\frac{\langle\NCS\rangle}{(L m_W)^3}
&=& (0.32 \pm 0.15) \times 10^{-5}\, k,
\;\;\;\; m_H=m_W,
\label{slopes2a}\\
&=& (0.79 \pm 0.36) \times 10^{-5}\, k,
\;\;\;\; m_H = \sqrt{2}\, m_W.
\label{slopes2}
\eea
This fit takes into account that with infinite statistics the result
at $k=0$ should be zero.
However, the large $\ch^2$ in the $m_H=m_W$ case suggests contamination
from non-linear behavior.

To diminish the uncertainty of where the linear regime ends we
focus on the lowest $k$-values,
$k=0,3$, $m_H=m_W$, and $k=0,1.5$, $m_H = \sqrt{2}\, m_W$.
We expect these points to be close to, or in, the linear regime
that is of physical relevance. Furthermore, since we used exactly
the same initial configurations for the data at
$k=0$ and at $k= 1.5, 3$, the effect of CP violation is the {\it
difference} between the zero and non-zero
$k$ results. An overall shift due to finite statistics
cancels out in this difference. These fits are represented by the
continuous lines in figure \ref{averages5}. The fitted slopes are
given by
\bea
\frac{\langle\NCS\rangle}{(L m_W)^3}
&=& (0.79 \pm 0.31) \times 10^{-5}\, k,
\;\;\;\; m_H=m_W,
\label{slopes1a}\\
&\simeq& 1.4  \times 10^{-5}\, k,
\;\;\;\; m_H = \sqrt{2}\, m_W.
\label{slopes1}
\eea
The error in the $m_H=m_W$ case is obtained by shifting the data
upward with the negative of the $k=0$ value and combining the
errors at $k=3$ in quadrature. For the case $m_H=\sqrt{2}\, m_H$
this would lead to a too large, not meaningful error, which we
have left out.  We consider as
our best estimate for the slope the result (\ref{slopes2})
from the previous fit (upper dashed line in figure \ref{averages5}),
keeping in mind that this is perhaps an under-estimate because
of the apparent curvature in the data.

\subsection{Initial rise}
\label{bump}
It is clearly difficult to predict the final averaged Chern-Simons
number analytically. For early times, specifically during the very
first roll-off of the Higgs field, we can however understand the
asymmetry generated by the driving force of the CP-violating term,
through an analysis similar to that used in the analog 1+1 D
Abelian Higgs model
\cite{Smit:2002yg}. Consider a spatial region in which the fields are
gauge-equivalent to being homogeneous and choose a gauge in which
they actually are homogeneous.
For homogeneous fields, the equations
of motion for the gauge field read:
\be
\partial_t^2 {A}_{k}^{a}=
A_{k}^{b}A_{l}^{b}A_{l}^{a}-A_{k}^{a}A_{l}^{b}A_{l}^{b} -
\frac{1}{2}\,g^2 A_{k}^{a}\phi^{\dagger}\phi
-2 g^2 \kappa\, B_{k}^{a}\,
\partial_t (\phi^{\dagger}\phi)
\ee
In the Abelian Higgs model in 1+1 dimensions \cite{Smit:2002yg},
the simplicity of the C-bias term ensures that the gauge field
does not enter into the driving force (the $\kappa$-term). In the
present case, however, its effect is suppressed by
\be
B_{k}^{a} = \half\,
\epsilon_{klm}\,\epsilon_{abc}A_{l}^{b}A_{m}^{c}.
\ee
To estimate the asymmetry, we assume that the Higgs field, which
may be viewed as an O(4) vector, rolls off the quadratic
potential, say
in the direction $\ph_3$:
\be
\label{pfield}
\partial_t^2 \phi_{3}=\mu^{2}\phi_{3}\rightarrow
\phi^{\dagger}\phi=\phi^{2}_{0}\, e^{2\mu t}.
\ee
Assume that we have solved the system for $\kappa=0$. From the
simulations, we know that the gauge field also grows
exponentially, say
\be
\label{bfield}
A_{k,0}^{a}= C_{k}^{a}e^{c\mu t}.
\ee
We perturb with the $\kappa$ term, and write
\be
A_{k}^{a}=A_{k,0}^{a}+\kappa A_{k,1}^{a}.
\ee
Then the equation for the perturbation $A_{k,1}^{a}$ is
\be
\partial_t^2 A^{a}_{k,1}=-H_{kl}^{ab}A_{l,1}^{b} -2 g^2
B_{k,0}^{a} \, \partial_t(\phi^{\dagger}\phi),
\label{kpterm}
\ee
with
\be
H^{ab}_{kl} = (A^2 \dl_{kl} - A^c_k A^c_l)\dl_{ab} - A^a_m A^b_m
\dl_{kl} + 2 A^a_k A^b_l - A^a_l A^b_k +
\half\, g^2 \phd\ph\, \dl_{kl}\dl_{ab},
\ee
evaluated at zeroth order in $\kp$. For small times $H$ is
dominated by the $\phd\ph$ term, which acts like an
effective-frequency term for the perturbation, that will turn the
initial rise induced by the source term $\propto
\partial_t(\phd\ph)$ into the small initial bump in figures
\ref{averages1} and
\ref{averages3}.
Assuming this effective mass term can be neglected at early times we can
integrate (\ref{kpterm}) using (\ref{pfield}) and (\ref{bfield})
and express the result as
\be
A^a_{k,1} = -  \frac{2\mu}{(2\mu + 2c\mu)^2}\,2 g^2\, B^a_{k,0}\,
\phd\ph.
\ee
For homogeneous fields the Chern-Simons density is given by
(cf.\ (\ref{CScurrent}))
\be
n_{\rm CS}(t) = \int_0^t dx^0\, \dnot j^0_{\rm CS} =
-\frac{1}{48\pi^2}\, \ep_{klm} \ep_{abc}\, A^a_k A^b_l A^c_m
\ee
evaluated at time $t$. Averaging over initial conditions the
zeroth order contribution in $\kp$ will vanish and we get
\be
\langle n_{\rm CS}\rangle = -\kp \frac{1}{8\pi^2}\, \langle
B^a_{k,0} A^a_{k,1}\rangle = \frac{\kp g^2}{8\pi^2 (1+c)^2
\mu}\,\langle B^2 \phd\ph\rangle.
\ee
We now assume this expression, which was derived for our
homogenous patch, to be independent of where the patch was
located, and replace
$\langle B^2 \phd\ph\rangle \to
\langle \overline{B^2} \rangle \langle \overline{\phd\ph}\rangle$,
where
$\langle \overline{B^2}\rangle$ and $\langle \overline{\phd\ph}\rangle$
stand for the average over the volume as well as over initial
conditions. This gives for the density in W-mass units,
\be
\frac{\langle n_{\rm CS}\rangle}{m_W^3} =
\frac{\sqrt{2}\, k}{(8\pi^2)^2 (1+c)^2}\,
\frac{\langle \overline{B^2}\rangle\langle \overline{\phd\ph}\rangle (2/v^2)}
{m_W^3 m_H}.
\label{rise}
\ee
Using the computed values of $\langle \overline{B^2} \rangle$ 
and $\langle \overline{\phd\ph} \rangle$, and inferring
$c$ from the measured behavior
$\langle \overline{B^2} \rangle \propto \exp(4c\mu t)$,
we can compare (\ref{rise}) to the data. 
For $c$ we find $c\simeq 0.67$.

\FIGURE{\epsfig{file=bumpkappa.eps,width=13cm,clip}
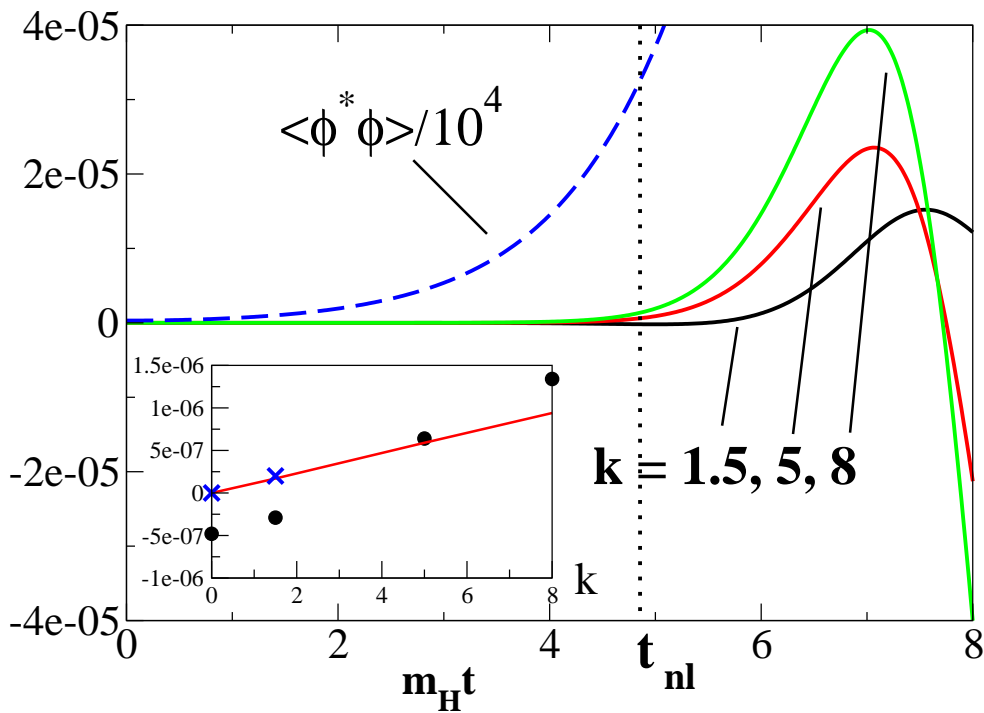
\caption{The initial rise for various $k$'s.
The inset shows the size of the bump versus $k$
for the estimate (\ref{rise}) (line), the
results of the simulations (dots) and results shifted by the negative
of the value at zero $k$ (see main text);
$m_{H}/m_{W}=\sqrt{2}$.
\label{bump1}}}

Fig.\ \ref{bump1} shows the initial bump for
$m_{H}/m_{W}=\sqrt{2}$, for varying $k$. The inset shows a
comparison with the estimate (\ref{rise}), evaluated at a time
for which $\langle\overline{\phi^{\dagger}\phi}\rangle= v^2/6$ ,
the value at which the curvature in the Higgs potential changes
sign and the quadratic approximation (\ref{pfield}) will be become
more strongly affected by non-linearities. Because of finite
statistics, there is an overall shift of the results that can be
read off from the $k=0$ simulation, which should give zero
with an infinite number of initial configurations. 
In this case the $k=0$ and $1.5$ simulations started from
the exact same initial conditions, and so we shift the
$k=0$ result to zero, and the $k=3$ result by the same amount
(as we did in section \ref{averages}). These are the crosses in
the inset. The agreement of (\ref{rise}) with the data is
surprisingly good, given the approximations made.

\subsection{Final temperature and sphaleron wash-out}
\label{temperature}
In order for baryogenesis to be successful, we need the final
temperature after symmetry breaking to be low enough so that an
asymmetry does not get washed out by equilibrium sphaleron
processes. The usual requirement for avoiding subsequent
wash-out, namely, that the sphaleron energy is much larger than
the temperature, can be conservatively formulated as \cite{Kajantie:1996kf}
$v/T > 1.49$, or $T/v = 0.67$.
The estimate (\ref{Testimate}), based on distributing the
Higgs potential energy ($V_{0} = v^2 m_H^2/8$) over $g_{*}$
relativistic degrees of freedom, leads to
\be
\frac{T}{v} = 0.442\,\left(\frac{m_H}{v}\right)^{1/2},
\ee
where we used $g_* = 10$ for the Higss and W degrees of freedom.
With $v=246$ GeV and a Higgs mass in the range 114 -- 200 GeV
\cite{Hagiwara:2002fs}, the requirement is
amply satisfied. For example, for $m_H =\sqrt{2}\, m_W\simeq 114$ GeV
the estimated temperature is $T=74 \, \mbox{GeV}$.

In fact, our system is not yet in equilibrium after the transition
and our particles are not massless. In \cite{Skullerud:2003ki} we
computed Higgs and W particle numbers using the field
configurations produced in our numerical simulation, from which we
obtained effective temperatures and chemical potentials by
comparing to a Bose--Einstein distribution. We found
\bea
\frac{T_{\rm eff}}
{m_{H}}\simeq 0.4,
\;\;\;\;
\frac{\mu^{\rm ch}_{\rm eff}}{m_{H}}\simeq 1,
\;\;\;\; m_H = \sqrt{2}\, m_W.
\eea
At first glance this temperature should keep us safe from
equilibrium sphaleron wash-out. The implied temperature
$0.4 \times 114 = 45$ GeV is lower than the 74 GeV found above,
which can be ascribed to the occurrence of the chemical potential.
However, this large chemical
potential implies that the occupation numbers are still very large
for the low momentum modes, $n_k = [\exp(\sqrt{m^2 + k^2}/T -
\mu^{\rm ch}/T)-1]^{-1} \gg 1$. Since these modes are the ones
responsible for sphaleron processes, the actual rate may be
larger than one would expect from the temperature alone.

In the Standard Model, the W and Higgs particles decay into the
lighter particles, which will rapidly lead to a lowering of the
temperature. For the W, the decay width is about 3 GeV and the
Higgs width is expected to be similar, so that in
a time span of a few hundred $\mhinv$ these particles have decayed. We
do not see any wash-out on time scales of a few hundred
$m_{H}^{-1}$. Then the final temperature will be determined by
light degrees of freedom and thus close to the estimate
(\ref{Testimate}) with the somewhat more favorable $g_* = 86.25$:
\be
\frac{T}{v} = 0.258\,\left(\frac{m_H}{v}\right)^{1/2}
\label{Testimate2}
\ee
(and a temperature $T=43$ GeV for $m_H = 114$ GeV).
We do not see a problem with wash-out.

\subsection{Effective sphaleron rate}

In \cite{Garcia-Bellido:1999sv} an effective sphaleron rate was introduced
and used in estimating the effect of the CP-bias term
using quasi-equilibrium concepts,
\be
n_B \approx \frac{\dl_{\rm CP}}{M^2}\,
\Gm_{\rm eff}\frac{\langle\phd\ph\rangle}{T_{\rm eff}},
\label{nbsphal}
\ee
where $\dl_{\rm CP}/M^2=16\pi^2\kp/3$ and $T_{\rm eff}$ is an
effective temperature estimated to be about 350 GeV. This
effective sphaleron rate was computed recently
\cite{Garcia-Bellido:2003wd},
with parameters different from ours (and with a finite quenching
rate), and it is interesting to see its behavior in our
simulation. The effective rate per unit volume can be defined as
\be
\Gm_{\rm eff} = \frac{1}{L^3}\, \frac{d}{dt}\left[
\langle \NCS(t)^2\rangle -
\langle\NCS(t)\rangle^2\right],
\label{sphalrate}
\ee
where we set $\NCS=0$ at $t=0$.
\FIGURE{\epsfig{file=sphaleronrate_k0.eps,width=13cm,clip}
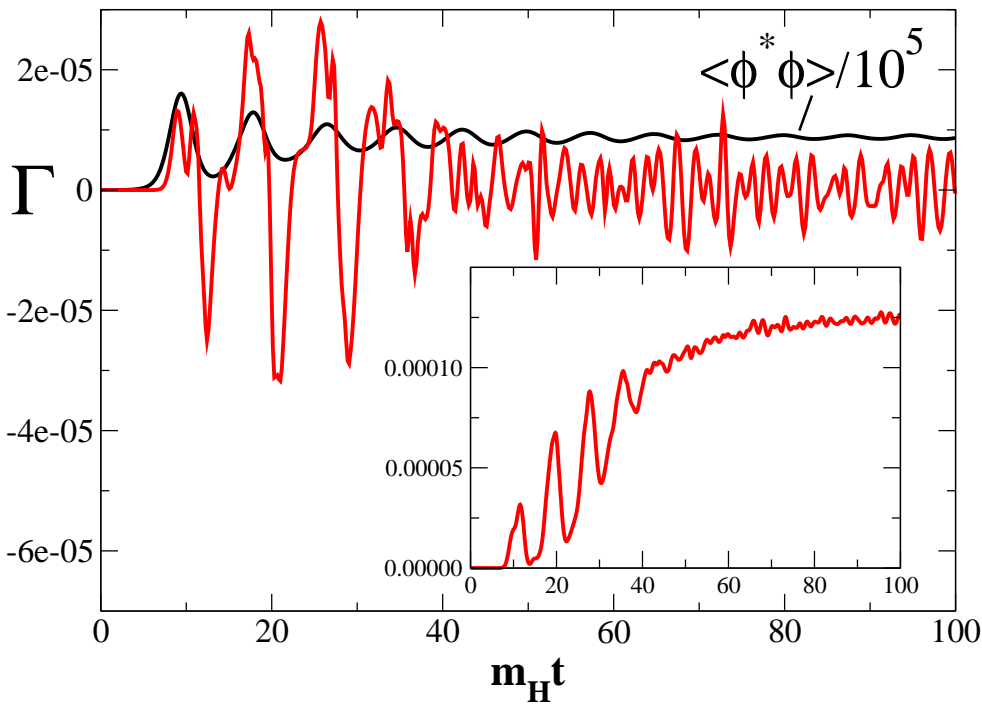
\caption{The effective sphaleron rate $\Gm_{\rm eff}/m_H^4$;
$m_{H}/m_{W}=1$, $k=0$, Thermal initial conditions.
Inset: integrated rate $\int_0^t dt'\,\Gm_{\rm eff}(t')/m_H^3$.
}\label{washout1}}
In Fig.\
\ref{washout1} we show $\Gm_{\rm eff}$ versus time. It turns out to be
insensitive to the choice of initial conditions, and insensitive
to $m_{H}/m_{W}$. In equilibrium this quantity is the diffusion
rate of Chern-Simons number, in which case it is the slope of a
straight line approximating $\langle \NCS^{2}\rangle-\langle
\NCS\rangle^{2}$ (see inset).

In the tachyonic transition such a diffusion regime is preceded by
a period of preheating, where the gauge fields acquire energy and
as a result a non-zero
$\langle \NCS^2\rangle$. The early-time rate at which
this happens does not have anything to do with the equilibrium
sphaleron rate. Rather, it is determined by the time scale of the
preheating mechanism. Note how this rate is initially in phase
with the Higgs field, but it has a large magnitude for small
{\it and large} values of
$\langle\phi^{\dagger}\phi\rangle$. At these early
times the system is not experiencing potential barriers
proportional to $\langle \phi^{\dagger}\phi\rangle$, with
sphalerons mediating Chern-Simons number change.
Even at the latest time the rate is not as small as the
temperature would suggest. Averaging out the oscillations in the
region $10 \lesssim t m_H \lesssim 40$, the effective slope (rate)
is about
$\Gamma_{\rm eff}/m_{H}^{4}\approx 0.35 \times 10^{-5}$, about the
same value as obtained in \cite{Garcia-Bellido:1999sv}.
By (\ref{nbsphal}) this leads to the estimate (for $m_H = m_W$)
\be
n_B/m_W^3 \approx 0.1 \times 10^{-5}\, k,
\ee
an order of magnitude smaller than our result (\ref{slopes1a})
(which should be multiplied by 3 to get $n_B/m_W^3$).

\section{Conclusion}
\label{Conc}
Using numerical simulations that include effective CP violation we
have obtained results for the Chern-Simons asymmetry as a function
of CP-violating parameter $\kp$. This can now be used to estimate
the value of
$\kp$ needed to reproduce the observed baryon asymmetry
$n_B/n_{\gm}$. Following the usual arguments, this ratio is
related to $n_B/s$, which is approximately conserved in time,
where $s$ is the entropy density. After BBN, $s=7.04\, n_{\gm}$
\cite{KT}, whereas after the tachyonic
electroweak transition it is given by
$s=\left(2\pi^2/45\right) g_{*}T^{3}$, $g_*=86.25$
the effective number of d.o.f.\ in quarks, leptons, gluons and
photons, and $T = 4.0\sqrt{m_H(\mbox{GeV})}$ GeV from
(\ref{Testimate2}). We also need
$m_W = 80.5$ GeV \cite{Hagiwara:2002fs}. Our results
(\ref{slopes2},\ref{slopes1a}) for the asymmetry ($k=16\pi^2
\kp\,m_W^2$) now lead to
\bea
\frac{n_{B}}{n_{\gamma}}&=& (0.79 \pm 0.31)\times 10^{-2}\kappa\,
m_{W}^{2}
\;\;\;\;\;\;\;
(m_{H}= m_{W}),\\
&=& (0.46\pm 0.21) \times 10^{-2}\kappa\, m_{W}^{2}
\;\;\;\;\;\;\;
(m_{H}=\sqrt{2}\, m_{W}).
\label{asym2}
\eea
Given our approximations of an instantaneous quench, initial
temperature zero and the neglect of damping effects from degrees
of freedom left out in the simulation, we consider these upper limits
on the generated baryon asymmetry\footnote{Note,
however, that we have also not taken into account the
effect of CP violation {\em before} the transition. This may
bias the initial distribution into an asymmetric form that still has
zero average baryon number. See also \cite{Nauta:2002ru,SEWM02BJN}.}
(the value (\ref{asym2}) is probably somewhat low, cf.\ sect.\ \ref{averages}).
To reproduce the observed
baryon asymmetry (\ref{ObsAsy}) for $m_H=\sqrt{2}\, m_W \simeq 114$ GeV,
$\kp$ has to be
\be
\kappa\simeq\frac{1.4 \times 10^{-7}}{m_{W}^{2}}
\simeq \frac{2.2\times 10^{-5}}{1{\rm TeV}^{2}},
\ee
which does not seem to be particularly large. Phrasing it
differently, venturing the scale
$M=m_W$ (cf.\ (\ref{deltacp})), it means
\be
\dl_{\rm CP} \simeq 0.7\times 10^{-5},
\ee
which is smaller than the Standard Model
$J \simeq 3\times 10^{-5}$ (cf.\ (\ref{J})). We see this as
encouragement for further pursuing scenarios for ESM electroweak
baryogenesis.

\acknowledgments
We thank Jelper Striet, Alejandro Arrizabalaga, Jon-Ivar Skullerud,
Bartjan van Tent and Leo van den Horn for useful discussions.
This work was supported in part by FOM/NWO.
AT enjoyed support from the ESF network COSLAB.

\appendix
\section{Lattice formulation}
This appendix contains details of the lattice formulation. The
lattice spacings in the three spatial directions are $a_1 =
a_2=a_3= a$, and in the time direction it is $a_0\equiv a_t$ (all
$a_{\mu}$ are positive). The parallel transporter, also called the link variable, from
lattice point $x+ a_{\mu}\hmu$ to $x$ is\footnote{In this appendix
we use summation convention for the group indices
$a,b,\ldots$, but not for the space-time indices $\mu,m,n,\ldots$.}
$U_{\mu,x} = U^{\dagger}_{-\mu,x+a_\mu\hmu} =
\exp[-ia_{\mu} A_{\mu}^b(x) \ta^b/2]$, such that e.g.\ the
forward covariant derivative on the Higgs doublet is
\be
D_{\mu} \ph(x) = [U_{\mu,x}\ph(x+a_{\mu} \hmu) - \ph(x)]/a_{\mu}.
\ee
In the following we use a re-scaled matrix form of the Higgs field
defined by
\be
\Phi_x= \sqrt{\lambda}\,a  \left[
\begin{array}{cc}
\phi_{2}^{*}(x) & \phi_{1}(x)\\
-\phi_{1}^{*}(x) & \phi_{2}(x)\\
\end{array}
\right],
\;\;\;\;
\frac{1}{2}\Tr[\Phi^{\dagger}\Phi]=
\lambda a^{2}\, \phi^{\dagger}\phi.
\ee
Furthermore, we shall use lattice units for $x$ and the covariant
derivatives, e.g.\
\be
D_{\mu} \Phi_x = U_{\mu,x}\Phi_{x+\hmu} -\Phi_x.
\ee
The backward covariant derivative (indicated with a prime) reads
\be
D'_{\mu} \Phi_x = \Phi_x - U^{\dagger}_{\mu, x-\hmu}
\Phi_{x-\hmu}= \Phi_x - U_{-\mu,x}\Phi_{x-\hmu}.
\ee

\subsection{Action}
We discretize the action on a space-time lattice of size
$L^3 t=(N a )^3\times N_{t}a_{t}$
with $a_{t}/a  \ll 1$ in the following way \cite{Ambjorn:1991pu}:
\bea
S_{\rm L}&=&\sum_{x}
\left[\beta_{G}^{t}\sum_{n}\left(1-\frac{1}{2}\Tr[U_{0n,x}]\right)
- \beta_{G}^s\sum_{n>m}\left(1-\frac{1}{2}\Tr
[U_{mn,x}]\right)\right.
\nonumber\\
&&\left.\mbox{}
+ \beta_{H}^{t}\frac{1}{2}\Tr
\left(D_{0}\Phi_{x}\right)^{\dagger}\left(D_{0}\Phi_{x}\right) -
\beta_{H}^s\sum_{n}\frac{1}{2}\Tr\left(D_{n}\Phi_{x}\right)^{\dagger}\left(D_{n}\Phi_{x}\right)\right.
\nonumber\\
&&\left.\mbox{}
- \beta_{R}\left(\frac{1}{2}\Tr[\Phi_{x}^{\dagger}\Phi_{x}]-v_{\rm lat}^{2}\right)^{2}
-\beta_{\kappa}\frac{1}{2}\Tr[\Phi_{x}^{\dagger}\Phi_{x}]
\ffd
\right],
\eea
where 
$U_{0n,x}$ and $U_{mn,x}$ are timelike and spacelike
plaquette fields defined as 
\be
U_{\mu\nu,x}=U_{\nu\mu,x}^{\dagger}=
U_{\mu,x}U_{\nu,x+\hmu}U^{\dagger}_{\mu,x+\hnu}U^{\dagger}_{\nu,x}.
\ee
The CP-violating term contains
\be
\ffd =
-\sum_{\mu\nu\sigma\rho}\epsilon^{\mu\nu\sigma\rho}\frac{1}{2}
\Tr[\bar{U}_{\mu\nu,x}\bar{U}_{\sigma\rho,x}],
\label{ffd}
\ee
where $\bar{U}_{\mu\nu,x}$ is a symmetrized plaquette field:
\be
\bar{U}_{\mu\nu,x}=
\frac{1}{4}(U_{\mu\nu,x}+U_{-\mu-\nu,x}+U_{-\nu\mu,x}+U_{\nu-\mu,x}).
\label{symplaq}
\ee
Matching continuum and lattice parameters we find:
\bea
&&\beta_{G}^{t}=\frac{4}{g^{2}}\frac{a}{a_{t}}
,~\beta_{G}^s=\frac{4}{g^{2}}\frac{a_{t}}{a}
,~\beta_{H}^{t}=\frac{1}{\lambda}\frac{a}{a_{t}}
,~\beta_{H}^s=\frac{1}{\lambda}\frac{a_{t}}{a},\\
&&\beta_{R}=\frac{1}{\lambda}\frac{a_{t}}{a}
,~\beta_{\kappa}=\frac{\kappa}{a^{2}\lambda}
,~v_{\rm lat}^{2}=\frac{(am_{H})^{2}}{4}.
\eea

\subsection{Equations of motion}
\label{eom}
Varying the action with respect to $A_{\mu,x}^a$ gives the
classical equations of motion:
\bea
\partial_{0}'\partial_{0}\Phi_{x}
&=&
\frac{\beta_{H}^s}{\beta_{H}^{t}}\sum_n D_{n}'D_{n}\Phi_{x}-
\frac{2\beta_{R}}{\beta_{H}^{t}}\left(\frac{1}{2}\Tr\left[\Phi_{x}^{\dagger}\Phi_{x}\right]-
\frac{a^2 m_{H}^{2}}{4}\right)\Phi_{x}
\nonumber\\&&
\mbox{}+ \frac{2\bt_{\kp}}{\bt^t_H} \ffd \Phi_x,
\label{higgs_eom}
\\
-\partial_{0}^{\prime}E_{n,x}^{a}
&=&
\frac{\beta_{G}^s}{\beta_{G}^{t}}\sum_m D_{m}^{\prime ab}\Tr\left[i\tau^{b}U_{mn,x}\right]+
\frac{2\beta_{H}^s}{\beta_{G}^{t}}\Tr\left[(D_{n}\Phi_{x})^{\dagger}i\tau^{a}\Phi_{x}\right]
\nonumber\\&&\mbox{}
+\frac{2\beta_{\kappa}}{\beta_{G}^{t}}\sum_{kl}\epsilon_{nkl}
\left\{\partial_{0}^{\prime }
\left(\Tr\left[i\ta^a U_{n,}U_{-l-k,x+\hat{n}+\hat{0}}U_{n,x+\hat{0}}^{\dagger}\right]
\Tr\left[\Phi_{x+\hat n+\hat{0}}^{\dagger}\Phi_{x+\hat n+\hat{0}}\right]\right)\right.
\nonumber\\&&\left.\mbox{}
+D_{l}^{\prime ab}\left(
\Tr\left[i\ta^b (U_{n,x}U_{l,x+\hat{n}} + U_{l,x}U_{n,x+\hat{l}})U_{-k-0,x+\hat{n}+\hat{l}}
(U_{n,x+\hat l}^{\dagger}U_{l,x}^{\dagger} +
U_{k,x+\hat{n}}^{\dagger}U_{n,x}^{\dagger})
\right]\right.\right.
\nonumber\\&&\left.\left. \times
\Tr\left[\Phi_{x+\hat n + \hat l}^{\dagger}\Phi_{x+\hat n + \hat l}\right]\right)\right\}.
\label{gauge_eom}
\eea
Above, we have defined the
{\em lattice $E$-field} $E^{a}_{n,x}$ through
\be
E_{n,x} = i\Tr\left[U_{n,x}U^{\dagger}_{n,x+\hat{0}}\ta^a\right].
\ee
We have chosen the temporal gauge,
$U_{0,x}=0$, and
$D_m^{\prime ab}$ is the backward covariant derivative in the adjoint
representation, e.g.\
\be
D_m^{\prime ab} E^b_{j,x} =
E^a_{j,x}-\half\Tr\left[U^{\dagger}_{-m,x}\ta^a
U_{-m,x}\ta^b\right] E^b_{j,x-\hat m}.
\ee
The variation with respect to $A_{0}$ gives the Gauss constraint:
\bea
0= C_{x}^{a}&\equiv&
-\sum_n D_{n}^{\prime ab}E_{x,n}^{b}
+\frac{2\beta_{H}^{t}}{\bt_G^t}
\Tr\left[\partial_{0}\Phi_{x}^{\dagger}i\tau^{a}\Phi_{x}\right]
\label{gauss_eom}\\&&\mbox{}
-\frac{2\beta_{\kappa}}{\beta_{G}^{t}}\sum_{jkl}\epsilon_{jkl}
D_{j}^{\prime ab}\left(\Tr[i\ta^b U_{j,x}U_{-l-k,x+
\hat{j}+\hat{0}}U_{j,x+\hat{0}}^{\dagger}
\Tr\left[\Phi_{x+\hat j+\hat 0}^{\dagger}\Phi_{x+\hat j+\hat
0}\right]\right).
\nonumber
\eea
A Noether argument tells us that the quantity $C^a_x$ is conserved
in (lattice) time for each $\vecx$ when evolving the system with
the equations of motion.

Because of the symmetrization in the discretization of $\Tr F\tilde F$,
which we found to be important for reducing discretization errors,
both the Higgs and the gauge field equations of motion are
{\it implicit}. They depend in a non-linear way on forward (in
time) links and Higgs fields. We solve these equations by
iteration. With sufficiently small time steps and/or $\kappa$
convergence should be good, and indeed we encountered no problems,
iterating between 3 and 9 times per time step with
$a_{t}/a=0.05$. However, the computational costs were
correspondingly large. We iterated to computer accuracy 
(double precision) and checked that the Gauss constraint is satisfied, again
to computer accuracy.

\subsection{Chern-Simons number and winding number}
The lattice expression for the change in Chern-Simons number is
given by
\be
\NCS(t)-\NCS(0) = \frac{1}{16\pi^2}\sum_0^t\sum_{\vecx} \ffd,
\ee
where $\ffd$ is given in (\ref{ffd}). The following lattice
implementation of the winding number in the Higgs field,
\be
N_{\rm w}=\frac{1}{192\pi^2}\sum_{\vecx,ijk}\epsilon_{ijk}
\Tr\left[\left(V_{x+\hat i}-V_{x-\hat i}\right)V_{x}^{\dagger}
\left(V_{x+\hat j}-V_{x-\hat j}\right)V_{x}^{\dagger}
\left(V_{x+\hat k}-V_{x-\hat k}\right)V_{x}^{\dagger}\right],
\ee
where $V_x=V(x)$ with $V(x)$ the SU(2) matrix defined in
(\ref{Vdef}), turned out to perform satisfactorily in our
simulation.

\section{Coping with the Gauss constraints}
\label{Gauss1}
\subsection{Initial Higgs fields}
Our task is to generate an ensemble of initial conditions for the
Higgs field in our classical simulations 
according to the distribution (\ref{Pinit},\ref{xidef},\ref{Pinit2}).
The distribution applies to a real scalar field with
the definitions
\be
\ph(\vecx) = \sum_{\veck} \frac{e^{i\veck\cdot\vecx}}{\sqrt{L^3}}\,
\ph_{\veck},
\;\;\;\;
\pi(\vecx) = \sum_{\veck} \frac{e^{i\veck\cdot\vecx}}{\sqrt{L^3}}\,
\pi_{\veck}.
\label{FouDef1}
\ee
In finite volume the momentum label takes on the discrete values
$\veck=2\pi \vecn/L$, and on the lattice with lattice spacing
$a=L/N$ we can choose $n_{j}=-(N/2-1), \ldots, 0, \ldots, N/2$.
The reality of the fields imposes
$\phi_{\veck}=\phi_{-\veck}^{\dagger}$ and similarly for the canonical
momenta, except for the special modes
$\vecn=(0 \;{\rm or}\, N/2,\, 0\; {\rm or}\, N/2,\, 0\; {\rm or}\,
N/2)$, for which $\exp(i\veck\cdot\vecx)$ is real, and so the
reality conditions imply that the corresponding
$\phi_{\veck}$ and $\pi_{\veck}$ are real. We shall term them ``corner'' modes.
There are $8$ of them. The rest we may as well name ``bulk''
modes. We can write these latter complex variables as
\be
\phi_{\veck}^{\alpha}=
\frac{1}{\sqrt{2\omega_{k}}}\left(a_{\veck}^{\alpha}+ib_{\veck}^{\alpha}\right),
\;\;\;\;
\pi_{\veck}^{\alpha}=
\sqrt{\frac{\omega_{k}}{2}}\left(c_{\veck}^{\alpha}+id_{\veck}^{\alpha}\right)
\label{FouDef2}.
\ee
We have put in a label
$\alpha=1,2,3,4$ for the real fields of the Higgs field
doublet:
\bea
\phi^{T}=\left(\frac{\phi^{1}+i\phi^{2}}{\sqrt{2}},
\frac{\phi^{3}+i\phi^{4}}{\sqrt{2}}\right),
\label{PhiDef}
\eea
and similar for the canonical conjugate $\pi$. 
On the lattice we have
\be
\omega_{k}\to \sqrt{\mu^2 + \veck_{\rm lat}^2},
\;\;\;\;
\veck_{\rm lat}^2 =
\sum_{j}\left(2-2\cos(k_{j}a)\right)/a^2,
\ee
and the lattice Higgs doublet is defined as
\be
\ph_\vecx = a \ph(x).
\ee
We continue to use lattice units and define the lattice canonical
conjugate of $\ph_\vecx$ as
\be
\pi_\vecx = \dnot\ph_\vecx = \ph_{\vecx + \hat 0} - \ph_\vecx.
\ee
For the corner modes, the
$b^{\alpha}_{\veck}$ and
$ d^{\alpha}_{\veck}$ are just zero.

For each spatial lattice site there are three Gauss constraints
(one for each generator of the gauge group) to be satisfied by the
initial configurations. In terms of\footnote{We denote the 
$2\times 2$ unit matrix by $\ta^4$.}
\be
\rh^{\al}_\vecx = ig^2 \half
\left(\pi^{\dagger}\tau^{\al}\phi-\phi^{\dagger}\tau^{\al}\pi\right)_\vecx
= i
\frac{2\beta_{H}^{t}}{\beta_{G}^{t}}\,\Tr[\dnot\Phi^{\dagger}\ta^{\al}\Phi]_{\vecx},
\;\;\;\;
\al=1, \dots, 4,
\ee
the constraints read
\be
\label{LocGauss}
\partial'_{n}E_{n,\vecx}^{a}= \rh^a_\vecx,
\ee
where we have used the fact that our initial
$A_{n}^{a}=0$. Integrating the left and right hand sides in a
periodic volume, this reduces to three global Gauss constraints,
\be
0 = \sum_{\vecx} \rh^{\bt}_\vecx,
\;\;\;\;
\bt=1,2,3.
\label{GauDef1}
\ee
These are three constraints on our many Fourier modes. Because we
have four real scalar fields it is convenient to impose and
additional fourth constraint:
\bea
0=\sum_\vecx \rho^{\bt}_\vecx,
\;\;\;\;
\bt=4,
\label{GauDef2}
\eea
which can be seen as the requirement that the total hypercharge
also vanishes in the periodic volume. Since it is only one extra
global constraint on the many modes we do not expect that imposing
it or not has a noticeable effect on the physical results.

Inserting equations (\ref{FouDef1}), (\ref{FouDef2}) and (\ref{PhiDef})
into (\ref{GauDef1}) and (\ref{GauDef2}) we get a set of constraints on
the $a_{\veck},b_{\veck},c_{\veck},d_{\veck}$, of which the last
one ($\bt=4$) is
\bea
0=\sum_{\alpha}\left[\sum_{\veck_{\rm
corner}}\left(c_{\veck}^{\alpha}a_{\veck}^{\alpha}\right)+
\sum_{\veck_{\rm bulk}}2\left(c_{\veck}^{\alpha}a_{\veck}^{\alpha}+d_{\veck}^{\alpha}b_{\veck}^{\alpha}\right)\right]
=\sum_{\alpha}\left[c_{\bf 0}^{\alpha}a_{\bf 0}^{\alpha}\right]
+f_{4},
\eea
where we have singled out the $\veck=(0,0,0)$ component for later
use, and dumped the rest into the quantity $f_{4}$. In a similar
way it is easy to find for the constraints for $\beta=1,2,3$:
\bea
0=c_{\bf 0}^{2}a_{\bf 0}^{3}-c_{\bf 0}^{3}a_{\bf 0}^{2}+c_{\bf 0}^{4}a_{\bf 0}^{1}-c_{\bf 0}^{1}a_{\bf 0}^{4}+f_{1},\\
0=c_{\bf 0}^{1}a_{\bf 0}^{3}-c_{\bf 0}^{3}a_{\bf 0}^{1}+c_{\bf 0}^{2}a_{\bf 0}^{4}-c_{\bf 0}^{4}a_{\bf 0}^{2}+f_{2},\\
0=c_{\bf 0}^{2}a_{\bf 0}^{1}-c_{\bf 0}^{1}a_{\bf 0}^{2}+c_{\bf
0}^{1}a_{\bf 0}^{4}-c_{\bf 0}^{4}a_{\bf 0}^{1}+f_{3},
\eea
where again $f_{1,2,3}$ are straightforward but long combinations
of the remaining variables.

These equations are linear in all the variables and can be written
as
\be
{\bf 0}={\bf M}{\bf c}_{\bf 0}+{\bf f},
\ee
with ${\bf M}={\bf M}({\bf a_{\bf 0}})$ a 4 by 4 matrix. We are
left with the task of generating sets of random numbers according
to the distribution
\be
P(a,b,c,d) \propto
\exp{\left(-\frac{1}{2}\sum_{\veck,\alpha}
\frac{a_{\veck}^{\alpha 2}+b_{\veck}^{\alpha 2}+c_{\veck}^{\alpha 2}+d_{\veck}^{\alpha 2}}
{n_{k}+1/2}\right)}\delta({\bf M}{\bf c_{0}}+{\bf f}).
\ee
We can integrate out the $c^{\alpha}_{\bf 0}$ to get
\be
P(a,b,c,d) \propto
\frac{1}{\det{\bf M}}\exp{\left(-\frac{1}{2}\sum_{\veck,\alpha}
\frac{\!\mbox{}^\prime\; a_{\veck}^{\alpha 2}+b_{\veck}^{\alpha 2}+c_{\veck}^{\alpha 2}+d_{\veck}^{\alpha 2}}
{n_{k}+1/2}-\frac{1}{2}\frac{{\bf G}^2}{n_{0}+1/2}\right)},
\ee
where the prime indicates that the sum over modes no longer
includes the $c_{\bf 0}^{\alpha}$, and
\bea
{\bf G}^{2}&=&({\bf M}^{-1}{\bf f})^{T}{\bf M}^{-1}{\bf f}=
\frac{\sum_{\alpha} f_{\alpha}^{2}}{\sum_{\alpha} a_{\bf 0}^{\alpha 2}},\\
\det{\bf M}&=&\sum_{\alpha} a_{\bf 0}^{\alpha 2}.
\eea
The distribution is no longer a product of Gaussians and we sample
it using a Monte-Carlo algorithm. For every realization of the
remaining unconstrained variables, we then solve for the $c_{\bf
0}^{\alpha}$, such that the global Gauss constraints are fulfilled.

Sampling a distribution with a $\delta$-function can be difficult,
and can require very elaborate Monte-Carlo algorithms. However,
having sufficiently many degrees of freedom to ``distribute the
constraints on'', the deviation from (in this case) a Gaussian
distribution is small. In our simulation, where there are about
50 unstable modes (when using Just-the-half) and the whole 
collection of lattice modes when using Thermal, we did not encounter
such problems.

\subsection{Initial gauge fields}
The Gauss constraint (eqs. \ref{gauss_eom}) imposes a constraint
on the gauge fields at each point in space, given a Higgs field
configuration. 
Keeping in mind that the gauge fields only grow large and
classical because of the coupling to the Higgs fields, we set the
gauge fields to zero initially. This simplifies the constraints
dramatically, to read:
\be
0=C_{x}^{a}=-\partial_{n}^{\prime}E_{n,x}^{a}+
\frac{2\beta_{H}^{t}}{\beta_{G}^{t}}
\Tr\left[\left(\partial_{0}\Phi_{x}\right)^{\dagger}i\ta^{a}\Phi_{x}\right].
\ee
We put $C_{x}^{a}$ to zero at each point in space, which means
that there are no sources other than the Higgs field. Given the
Higgs fields we can now write:
\be
\partial'_{n} E_{n,\vecx}^{a}(x)=\rho^{a}_\vecx,
\;\;\;\;
a=1,2,3,
\ee
which is just a Coulomb equation, and is solved by introducing a
set of scalar functions $\chi^{a}_\vecx$ so that
\bea
E_{n,\vecx}^{a}=-\partial_{n}\chi^{a}_\vecx.
\eea
Fourier transforming
\be
-\partial_{n}\partial_{n}^{\prime}\chi_\vecx^{a}=\rho^{a}_\vecx
\ee
gives
\bea
\tilde{\chi}^{a}(\veck)=
\frac{\tilde{\rho}^{a}_\veck}{\veck_{\rm lat}^{2}}.
\eea
Since we use a periodic volume, $\tilde\rh^a_{\veck}$ should
vanish at $\veck=0$, i.e.\ the global Gauss constraints on the
Higgs fields
\be
0= \sum_{\vecx} \partial_{n}^{\prime}E_{n,\vecx}^{a}=
\sum_\vecx \rho^{a}_\vecx
\ee
have to be satisfied. Their implementation is described in the
previous section.

\bibliography{lit}

\providecommand{\href}[2]{#2}\begingroup\raggedright\begin{thebibliography}{10}

\bibitem{Spergel:2003cb}
D.~N. Spergel {\em et.~al.}, {\it First year {W}ilkinson {M}icrowave
  {A}nisotropy {P}robe ({WMAP}) observations: Determination of cosmological
  parameters},  {\em Astrophys. J. Suppl.} {\bf 148} (2003) 175,
  [\href{http://xxx.lanl.gov/abs/astro-ph/0302209}{{\tt astro-ph/0302209}}].

\bibitem{Sakharov:1967dj}
A.~D. Sakharov, {\it Violation of {CP} invariance, {C} asymmetry, and baryon
  asymmetry of the universe},  {\em Pisma Zh. Eksp. Teor. Fiz.} {\bf 5} (1967)
  32--35.

\bibitem{Dine:2003ax}
M.~Dine and A.~Kusenko, {\it The origin of the matter-antimatter asymmetry},
  \href{http://xxx.lanl.gov/abs/hep-ph/0303065}{{\tt hep-ph/0303065}}.

\bibitem{'tHooft:1976up}
G.~'t~Hooft, {\it Symmetry breaking through {B}ell-{J}ackiw anomalies},  {\em
  Phys. Rev. Lett.} {\bf 37} (1976) 8--11.

\bibitem{Cabibbo:1963yz}
N.~Cabibbo, {\it Unitary symmetry and leptonic decays},  {\em Phys. Rev. Lett.}
  {\bf 10} (1963) 531--532.

\bibitem{Kobayashi:1973fv}
M.~Kobayashi and T.~Maskawa, {\it {CP} violation in the renormalizable theory
  of weak interaction},  {\em Prog. Theor. Phys.} {\bf 49} (1973) 652--657.

\bibitem{Hagiwara:2002fs}
{\bf Particle Data Group} Collaboration, K.~Hagiwara {\em et.~al.}, {\it Review
  of particle physics},  {\em Phys. Rev.} {\bf D66} (2002) 010001.

\bibitem{Buchmuller:2003gz}
W.~Buchmuller, P.~Di~Bari, and M.~Plumacher, {\it The neutrino mass window for
  baryogenesis},  {\em Nucl. Phys.} {\bf B665} (2003) 445--468,
  [\href{http://xxx.lanl.gov/abs/hep-ph/0302092}{{\tt hep-ph/0302092}}].

\bibitem{Kuzmin:1985mm}
V.~A. Kuzmin, V.~A. Rubakov, and M.~E. Shaposhnikov, {\it On the anomalous
  electroweak baryon number nonconservation in the early universe},  {\em Phys.
  Lett.} {\bf B155} (1985) 36.

\bibitem{Rubakov:1996vz}
V.~A. Rubakov and M.~E. Shaposhnikov, {\it Electroweak baryon number
  non-conservation in the early universe and in high-energy collisions},  {\em
  Usp. Fiz. Nauk} {\bf 166} (1996) 493--537,
  [\href{http://xxx.lanl.gov/abs/hep-ph/9603208}{{\tt hep-ph/9603208}}].

\bibitem{Cohen:1993nk}
A.~G. Cohen, D.~B. Kaplan, and A.~E. Nelson, {\it Progress in electroweak
  baryogenesis},  {\em Ann. Rev. Nucl. Part. Sci.} {\bf 43} (1993) 27--70,
  [\href{http://xxx.lanl.gov/abs/hep-ph/9302210}{{\tt hep-ph/9302210}}].

\bibitem{Kajantie:1996kf}
K.~Kajantie, M.~Laine, K.~Rummukainen, and M.~E. Shaposhnikov, {\it The
  electroweak phase transition: A non-perturbative analysis},  {\em Nucl.
  Phys.} {\bf B466} (1996) 189--258,
  [\href{http://xxx.lanl.gov/abs/hep-lat/9510020}{{\tt hep-lat/9510020}}].

\bibitem{Csikor:1998eu}
F.~Csikor, Z.~Fodor, and J.~Heitger, {\it Endpoint of the hot electroweak phase
  transition},  {\em Phys. Rev. Lett.} {\bf 82} (1999) 21--24,
  [\href{http://xxx.lanl.gov/abs/hep-ph/9809291}{{\tt hep-ph/9809291}}].

\bibitem{Carena:1996wj}
M.~Carena, M.~Quiros, and C.~E.~M. Wagner, {\it Opening the window for
  electroweak baryogenesis},  {\em Phys. Lett.} {\bf B380} (1996) 81--91,
  [\href{http://xxx.lanl.gov/abs/hep-ph/9603420}{{\tt hep-ph/9603420}}].

\bibitem{Laine:1998qk}
M.~Laine and K.~Rummukainen, {\it The {MSSM} electroweak phase transition on
  the lattice},  {\em Nucl. Phys.} {\bf B535} (1998) 423--457,
  [\href{http://xxx.lanl.gov/abs/hep-lat/9804019}{{\tt hep-lat/9804019}}].

\bibitem{Garcia-Bellido:1999sv}
J.~Garcia-Bellido, D.~Y. Grigoriev, A.~Kusenko, and M.~E. Shaposhnikov, {\it
  Non-equilibrium electroweak baryogenesis from preheating after inflation},
  {\em Phys. Rev.} {\bf D60} (1999) 123504,
  [\href{http://xxx.lanl.gov/abs/hep-ph/9902449}{{\tt hep-ph/9902449}}].

\bibitem{Krauss:1999ng}
L.~M. Krauss and M.~Trodden, {\it Baryogenesis below the electroweak scale},
  {\em Phys. Rev. Lett.} {\bf 83} (1999) 1502--1505,
  [\href{http://xxx.lanl.gov/abs/hep-ph/9902420}{{\tt hep-ph/9902420}}].

\bibitem{Copeland:2001qw}
E.~J. Copeland, D.~Lyth, A.~Rajantie, and M.~Trodden, {\it Hybrid inflation and
  baryogenesis at the {TeV} scale},  {\em Phys. Rev.} {\bf D64} (2001) 043506,
  [\href{http://xxx.lanl.gov/abs/hep-ph/0103231}{{\tt hep-ph/0103231}}].

\bibitem{Rajantie:2000nj}
A.~Rajantie, P.~M. Saffin, and E.~J. Copeland, {\it Electroweak preheating on a
  lattice},  {\em Phys. Rev.} {\bf D63} (2001) 123512,
  [\href{http://xxx.lanl.gov/abs/hep-ph/0012097}{{\tt hep-ph/0012097}}].

\bibitem{Copeland:2002ku}
E.~J. Copeland, S.~Pascoli, and A.~Rajantie, {\it Dynamics of tachyonic
  preheating after hybrid inflation},  {\em Phys. Rev.} {\bf D65} (2002)
  103517, [\href{http://xxx.lanl.gov/abs/hep-ph/0202031}{{\tt
  hep-ph/0202031}}].

\bibitem{Garcia-Bellido:2002aj}
J.~Garcia-Bellido, M.~Garcia~Perez, and A.~Gonzalez-Arroyo, {\it Symmetry
  breaking and false vacuum decay after hybrid inflation},  {\em Phys. Rev.}
  {\bf D67} (2003) 103501, [\href{http://xxx.lanl.gov/abs/hep-ph/0208228}{{\tt
  hep-ph/0208228}}].

\bibitem{Smit:2002yg}
J.~Smit and A.~Tranberg, {\it Chern-{S}imons number asymmetry from {CP}
  violation at electroweak tachyonic preheating},  {\em JHEP} {\bf 12} (2002)
  020, [\href{http://xxx.lanl.gov/abs/hep-ph/0211243}{{\tt hep-ph/0211243}}].

\bibitem{Garcia-Bellido:2003wd}
J.~Garc\'{\i}a-Bellido, M.~Garcia-Perez, and A.~Gonzalez-Arroyo, {\it
  Chern-{S}imons production during preheating in hybrid inflation models},
  \href{http://xxx.lanl.gov/abs/hep-ph/0304285}{{\tt hep-ph/0304285}}.

\bibitem{Jarlskog:1985ht}
C.~Jarlskog, {\it Commutator of the quark mass matrices in the standard
  electroweak model and a measure of maximal {CP} violation},  {\em Phys. Rev.
  Lett.} {\bf 55} (1985) 1039.

\bibitem{Shaposhnikov:1987tw}
M.~E. Shaposhnikov, {\it Baryon asymmetry of the universe in standard
  electroweak theory},  {\em Nucl. Phys.} {\bf B287} (1987) 757--775.

\bibitem{Shaposhnikov:1988pf}
M.~E. Shaposhnikov, {\it Structure of the high temperature gauge ground state
  and electroweak production of the baryon asymmetry},  {\em Nucl. Phys.} {\bf
  B299} (1988) 797.

\bibitem{Farrar:1993sp}
G.~R. Farrar and M.~E. Shaposhnikov, {\it Baryon asymmetry of the universe in
  the minimal {S}tandard {M}odel},  {\em Phys. Rev. Lett.} {\bf 70} (1993)
  2833--2836, [\href{http://xxx.lanl.gov/abs/hep-ph/9305274}{{\tt
  hep-ph/9305274}}]. [Erratum-ibid.71:210,1993].

\bibitem{Farrar:1994hn}
G.~R. Farrar and M.~E. Shaposhnikov, {\it Baryon asymmetry of the universe in
  the standard electroweak theory},  {\em Phys. Rev.} {\bf D50} (1994) 774,
  [\href{http://xxx.lanl.gov/abs/hep-ph/9305275}{{\tt hep-ph/9305275}}].
  [Erratum-ibid.71:210,1993].

\bibitem{Farrar:1994kf}
G.~R. Farrar and M.~E. Shaposhnikov, {\it Note added to '{B}aryon asymmetry of
  the universe in the standard model'},
  \href{http://xxx.lanl.gov/abs/hep-ph/9406387}{{\tt hep-ph/9406387}}.

\bibitem{Konstandin:2003dx}
T.~Konstandin, T.~Prokopec, and M.~G. Schmidt, {\it Axial currents from {CKM}
  matrix {CP} violation and electroweak baryogenesis},
  \href{http://xxx.lanl.gov/abs/hep-ph/0309291}{{\tt hep-ph/0309291}}.

\bibitem{Cosmo03JS}
J.~Smit, ``Tachyonic electroweak transition, {CP} violation and baryogenesis.''
\newblock Talk at Cosmo-03, Ambleside, UK.

\bibitem{Bodeker:1999gx}
D.~Bodeker, G.~D. Moore, and K.~Rummukainen, {\it Chern-{S}imons number
  diffusion and hard thermal loops on the lattice},  {\em Phys. Rev.} {\bf D61}
  (2000) 056003, [\href{http://xxx.lanl.gov/abs/hep-ph/9907545}{{\tt
  hep-ph/9907545}}].

\bibitem{Moore:1998sw}
G.~D. Moore, {\it Measuring the broken phase sphaleron rate nonperturbatively},
   {\em Phys. Rev.} {\bf D59} (1999) 014503,
  [\href{http://xxx.lanl.gov/abs/hep-ph/9805264}{{\tt hep-ph/9805264}}].

\bibitem{Skullerud:2003ki}
J.-I. Skullerud, J.~Smit, and A.~Tranberg, {\it W and {H}iggs particle
  distributions during electroweak tachyonic preheating},  {\em JHEP} {\bf 08}
  (2003) 045, [\href{http://xxx.lanl.gov/abs/hep-ph/0307094}{{\tt
  hep-ph/0307094}}].

\bibitem{Ambjorn:1991pu}
J.~Ambjorn, T.~Askgaard, H.~Porter, and M.~E. Shaposhnikov, {\it Sphaleron
  transitions and baryon asymmetry: A numerical real time analysis},  {\em
  Nucl. Phys.} {\bf B353} (1991) 346--378.

\bibitem{Grigoriev:1992nv}
D.~Y. Grigoriev, M.~E. Shaposhnikov, and N.~Turok, {\it Electroweak
  baryogenesis: A numerical study in (1+1)- dimensions},  {\em Phys. Lett.}
  {\bf B275} (1992) 395--402.

\bibitem{SEWM02AT}
J.~Smit and A.~Tranberg, {\it Chern-{S}imons number asymmetry from
  {CP}-violation during tachyonic preheating},  in {\em Strong and
  {E}lectroweak {M}atter 2002 -- {P}roceedings of the {SEWM}2002 {M}eeting}
  (M.~G. Schmidt, ed.), (Singapore), World Scientific, 2002.
\newblock
  \href{http://xxx.lanl.gov/abs/http://arXiv.org/abs/hep-ph/0210348}{{\tt
  http://arXiv.org/abs/hep-ph/0210348}}.

\bibitem{Smit:2003my}
J.~Smit and A.~Tranberg, {\it Classical issues in electroweak baryogenesis},
  \href{http://xxx.lanl.gov/abs/hep-lat/0309082}{{\tt hep-lat/0309082}}.

\bibitem{Cosmo03AT}
A.~Tranberg, ``Baryogenesys from {CP} violation at electroweak tachyonic
  preheating.''
\newblock Talk at Cosmo-03, Ambleside, UK.

\bibitem{KT}
E.~W. Kolb and M.~S. Turner, {\em The {E}arly {U}niverse}.
\newblock Addison-Wesley, Reading, {M}assachusetts, 1990.

\bibitem{Peebles:1998qn}
P.~J.~E. Peebles and A.~Vilenkin, {\it Quintessential inflation},  {\em Phys.
  Rev.} {\bf D59} (1999) 063505,
  [\href{http://xxx.lanl.gov/abs/astro-ph/9810509}{{\tt astro-ph/9810509}}].

\bibitem{German:2001tz}
G.~German, G.~Ross, and S.~Sarkar, {\it Low-scale inflation},  {\em Nucl.
  Phys.} {\bf B608} (2001) 423--450,
  [\href{http://xxx.lanl.gov/abs/hep-ph/0103243}{{\tt hep-ph/0103243}}].

\bibitem{Berera:2003kg}
A.~Berera and R.~O. Ramos, {\it Absence of isentropic expansion in various
  inflation models},  \href{http://xxx.lanl.gov/abs/hep-ph/0308211}{{\tt
  hep-ph/0308211}}.

\bibitem{Asaka:2001ez}
T.~Asaka, W.~Buchmuller, and L.~Covi, {\it False vacuum decay after inflation},
   {\em Phys. Lett.} {\bf B510} (2001) 271--276,
  [\href{http://xxx.lanl.gov/abs/hep-ph/0104037}{{\tt hep-ph/0104037}}].

\bibitem{Felder:2000hj}
G.~N. Felder {\em et.~al.}, {\it Dynamics of symmetry breaking and tachyonic
  preheating},  {\em Phys. Rev. Lett.} {\bf 87} (2001) 011601,
  [\href{http://xxx.lanl.gov/abs/hep-ph/0012142}{{\tt hep-ph/0012142}}].

\bibitem{Smit:2001qa}
J.~Smit, J.~C. Vink, and M.~Salle, {\it Initial conditions for simulated
  'tachyonic preheating' and the {H}artree ensemble approximation},
  \href{http://xxx.lanl.gov/abs/hep-ph/0112057}{{\tt hep-ph/0112057}}.
  Contribution to Cosmo-01, Rovaniemi, Finland.

\bibitem{Nauta:2002ru}
B.-J. Nauta and A.~Arrizabalaga, {\it Asymmetric {C}hern-{S}imons number
  diffusion from {CP}-violation},  {\em Nucl. Phys.} {\bf B635} (2002)
  255--285, [\href{http://xxx.lanl.gov/abs/hep-ph/0202115}{{\tt
  hep-ph/0202115}}].

\bibitem{SEWM02BJN}
B.-J. Nauta, {\it On asymmetric {C}hern-{S}imons number diffusion},  in {\em
  Strong and {E}lectroweak {M}atter 2002 -- {P}roceedings of the {SEWM}2002
  {M}eeting} (M.~G. Schmidt, ed.), (Singapore), World Scientific, 2002.
\newblock
  \href{http://xxx.lanl.gov/abs/http://arXiv.org/abs/hep-ph/0301150}{{\tt
  http://arXiv.org/abs/hep-ph/0301150}}.

\end{thebibliography}\endgroup
\end{document}